\documentclass[12pt]{article}

\pdfoutput=1

\usepackage{amsmath,amssymb}
\usepackage{bm}
\usepackage{graphicx, color}
\usepackage{wrapfig}
\begin{document}
\title{
\begin{flushright}
\ \\*[-80pt] 
\begin{minipage}{0.2\linewidth}
\normalsize
HUPD1801 \\*[50pt]
\end{minipage}
\end{flushright}
{\Large \bf 
 Revisiting $A_4$ model for leptons  in  light of NuFIT 3.2
\\*[20pt]}}

\author{ 
\centerline{
Sin~Kyu~Kang$^{1,}$\footnote{E-mail address: skkang@seoultech.ac.kr},
~Yusuke~Shimizu$^{2,}$\footnote{E-mail address: yu-shimizu@hiroshima-u.ac.jp},
~Kenta~Takagi$^{2,}$\footnote{E-mail address: takagi-kenta@hiroshima-u.ac.jp},} \\*[5pt]
\centerline{Shunya~Takahashi$^{2,}$\footnote{E-mail address: s-takahashi@hiroshima-u.ac.jp}, 
~Morimitsu~Tanimoto$^{3,}$\footnote{E-mail address: tanimoto@muse.sc.niigata-u.ac.jp}} 
\\*[20pt]
\centerline{
\begin{minipage}{\linewidth}
\begin{center}
$^1${\it \normalsize
School of Liberal Arts, Seoul-Tech, Seoul 139-743, Korea} \\*[5pt]
$^2${\it \normalsize
Graduate School of Science, Hiroshima University, Higashi-Hiroshima 739-8526, Japan} \\*[5pt]
$^3${\it \normalsize
Department of Physics, Niigata University, Niigata 950-2181, Japan}
\end{center}
\end{minipage}}
\\*[50pt]}

\date{
\centerline{\small \bf Abstract}
\begin{minipage}{0.9\linewidth}
\medskip 
\medskip 
\small 
We revisit the $A_4$ model for leptons in light of new result of NuFIT 3.2.
We introduce a new flavon $\eta$ transforming as $A_4$ singlet $1'$ or $1''$ which couples to both charged leptons and neutrinos in next-leading order operators.
The model consists of the five parameters:
the lightest neutrino mass $m_1$,
the vacuum expectation value of  $\eta$ and three CP violating phases
after inputting the experimental values of $\Delta m_{\rm atm}^2$ and 
$\Delta m_{\rm sol}^2$.
The model with the $1''$ singlet flavon gives the prediction of $\sin^2 \theta_{12}$ around the best fit of NuFIT  3.2
while keeping near the maximal mixing of $\theta_{23}$.
Inputting the experimental mixing angles with the $1\,\sigma$ error-bar, 
the Dirac CP violating phase is clearly predicted to be $|\delta_\text{CP}|=50^\circ- 120^\circ$, which will be tested by the precise observed value in the future.
 In order to get the best fit value $\sin^2\theta_{23}=0.538$,
the sum of three neutrino masses is predicted to be larger than $90$\,meV.
The cosmological observation for the sum of neutrino masses
will also provide a crucial test of our predictions.
It is remarked that the model is
consistent with the experimental data 
only for the normal hierarchy of neutrino masses.
\end{minipage}
}

\begin{titlepage}
\maketitle
\thispagestyle{empty}
\end{titlepage}

\section{Introduction}

The origin of the quark/lepton flavor is still unknown
in spite of the remarkable success of the standard model (SM).
To reveal the underlying physics of flavors  is the challenging work.
The recent development of the neutrino oscillation experiments provides us  with
 important clues to investigate the flavor physics.
Indeed,
the neutrino oscillation experiments
have determined precisely two neutrino mass squared differences and  three neutrino  mixing angles. 
Especially, the recent data of both T2K~\cite{Abe:2017vif, T2K}
and NO$\nu$A~\cite{Adamson:2017gxd, nova} give us that the atmospheric neutrino mixing angle $\theta_{23}$ favors near the  maximal angle $45^\circ$. The global analysis by NuFIT 3.2
presents the best fit $\theta_{23}=47.2^\circ$ for the normal hierarchy (NH) of neutrino masses~\cite{NuFIT}.
The closer the observed $\theta_{23}$ is to the maximal mixing, the more likely we are to believe in some flavor symmetry behind it.
In addition to the precise measurements of the mixing angles,
the T2K and NO$\nu$A strongly indicate the CP violation in the neutrino oscillation~\cite{T2K,nova}.
Thus, we are in the era to develop the flavor structure of the lepton mass matrices with focus on the leptonic flavor mixing angles and CP violating phase.

Before the reactor experiments measured non-zero value of $\theta_{13}$ in 2012\,\cite{An:2012eh,Ahn:2012nd}, the paradigm of the tri-bimaximal (TBM) mixing\,\cite{Harrison:2002er,Harrison:2002kp},
a highly symmetric mixing pattern for leptons, has attracted much attention.
It is well known that this mixing pattern is derived in the framework of the $A_4$ flavor symmetry
  \cite{Ma:2001dn}-\cite{Altarelli:2005yx}.
Therefore, non-Abelian discrete groups have become the center of attention in the flavor symmetry~\cite{Altarelli:2010gt}-\cite{King:2014nza}.
In order to obtain non-vanishing $\theta_{13}$,  two of the authors improved the $A_4$ model 
by the minimal modification through introducing another flavon which transforms as $1'('')$ of $A_4$
and couples only to the neutrino sector~\cite{Shimizu:2011xg}.
 Then, the predicted  values of $\theta_{13}$ are consistent with the experimental data.
 This pattern is essentially  the trimaximal mixing $\rm TM_2$~ \cite{Grimus:2008tt,Albright:2010ap,Rodejohann:2012cf} 
 which leads to $\sin^2 \theta_{12}\geq 1/3$.
 However, the predicted $\sin^2 \theta_{12}$ is outside of $2\,\sigma$ interval of the experimental data in the NuFIT 3.2 result~\cite{NuFIT}.
 Therefore,  the $A_4$ model should be reconsidered in  light of the new data of T2K and NO$\nu$A
as the implication of the new data has been changed.
 
 In this work, we introduce a new flavon transforming as $A_4$ singlet, $\eta$ ($1'$ or $1''$) which couples 
 to both charged leptons and neutrinos in next-leading order operators.
 The model consists of the five parameters:
 the lightest neutrino mass $m_1$,
 the vacuum expectation value (VEV) of  $\eta$ and three CP violating phases
 after inputting the observed values of $\Delta m_{\rm atm}^2$ and $\Delta m_{\rm sol}^2$.
 The model with a $1''$ singlet flavon gives the prediction of $\sin^2 \theta_{12}$ around the best fit of NuFIT  3.2
 with keeping near the maximal mixing of  $\theta_{23}$.
 The non-vanishing $\theta_{13}$ is derived from both charged lepton and neutrino sectors. 
 Inputting the observed mixing angles with the $1\,\sigma$ error-bar, 
 the CP violating Dirac phase is clearly predicted to be $|\delta_\text{CP}|=50^\circ- 120^\circ$.
 Therefore, the observation of the CP violating phase is essential to test the model in the future.
 
 It is  remarked that the model is consistent with the experimental data only for NH of neutrino masses.  
 The inverted hierarchy (IH) of neutrino masses is not allowed in the recent experimental data.
 This situation comes from that the singlet $1'$ or $1''$ flavon couples to leptons in the next-leading order.
 It is contrast with the model in Ref.\cite{Shimizu:2011xg} where both NH and IH are allowed.


We present our framework of the  $A_4$ model in section 2 where lepton mass matrices and VEVs of scalars are discussed.
The numerical results are shown in section 3. 
The section 4 is devoted to the summary.
Appendix A shows the lepton mixing matrix and  CP violating measures which are used in this work.
 The relevant  multiplication rules of $A_4$ are represented in  Appendix B.
The derivation of the lepton mixing matrix is given in Appendix C.
Appendix D presents the distributions of our parameter which are used in our numerical calculations.  

\vskip 1 cm
\section{Our framework of $A_4$ model}
We discuss  our $A_4$ model in the framework of the supersymmetry (SUSY). 
In the non-Abelian finite group $A_4$, there are four irreducible representations: 
$1$, $1'$, $1''$ and $3$.  
The left-handed leptons $l$ and right-handed charged leptons $e^c$, $\mu^c$, $\tau^c$ 
are assigned to the triplet and singlets, respectively, as seen in Table 1. 
The two Higgs doublets $(h_u, h_d)$ are assigned to the $A_4$ singlets, 
and their VEVs are denoted as $(v_u, v_d)$ as usual. 
We introduce several flavons as listed in Table 1.  
The flavons $\phi_T$ and  $\phi_S$ are $A_4$ triplets while $\xi$ and $\tilde\xi$ are the same  $A_4$ singlet $1$. 
In addition, $\eta$ and $\tilde\eta$ are the same non-trivial singlet $1''$ or $1'$. 
The $A_4$ flavor symmetry is spontaneously broken by VEVs of gauge singlet flavons, $\phi_T$, $\phi_S$, $\xi$ and  $\eta$,
whereas $\tilde \xi \ (1)$ and $\tilde\eta\ (1'',1')$ are defined to have vanishing VEVs
through the linear combinations of $\xi$ and $\tilde \xi$ and $\eta$ and $\tilde\eta$, respectively,
as discussed in Ref.~\cite{Altarelli:2005yx}.
In the original model proposed by Altarelli and Feruglio~\cite{Altarelli:2005yp, Altarelli:2005yx},
$\phi_T$, $\phi_S$ and $\xi$ were introduced, and then
the specific vacuum alignments of the triplet flavons lead to the tri-bimaximal mixing 
where the lepton mixing angle  $\theta_{13}$ vanishes.
In 2011, two of the authors minimally modified the model by introducing an extra flavon $\eta\ (1')$
on top of those flavons to generate non-vanishing $\theta_{13}$ \cite{Shimizu:2011xg}.
This modification of the model leads to the trimaximal mixing of neutrino flavors,
so called ${\rm TM_2}$ which predicts $\sin^2 \theta_{12}\geq 1/3$ \cite{Grimus:2008tt,Albright:2010ap,Rodejohann:2012cf}. Unfortunately, this prediction for $\theta_{12}$ is inconsistent with the data at $2\,\sigma$ C.L.
given in the NuFIT 3.2 result~\cite{NuFIT}.
In this work,  we force the flavon $\eta\ (1''\ {\rm or}\ 1')$  to couple
to both charged  lepton  and neutrino sectors in next-leading operators by assigning a $Z_3$ charge to $\eta$ appropriately.

We impose the $Z_3$ symmetry  to control Yukawa couplings in both neutrino sector and charged lepton sector.
The third row of Table 1 shows how each chiral multiplet transforms under $Z_3$ with its charge $\omega={\rm exp}(2\pi i/3)$.

 In order to obtain the natural hierarchy among lepton masses $m_e$, $m_\mu$ and $m_\tau$, we resort to the Froggatt-Nielsen mechanism \cite{Froggatt:1978nt} with an additional $U(1)_{\rm FN}$ symmetry under which only the right-handed lepton sector is charged. The field $\Theta$ denotes the Froggatt-Nielsen flavon in Table 1. The $U(1)_{\rm FN}$ charges are taken as ($4$, $2$, $0$) for ($e^c$, $\mu^c$, $\tau^c$), respectively.
 By assuming that $\Theta$, carrying a negative unit charge of $U(1)_{\rm FN}$, acquires a VEV, the relevant mass ratio is reproduced through the Froggatt-Nielsen charges.

 We also introduce a $U(1)_R$ symmetry in Table 1 to distinguish the flavons and driving fields $\phi^T_0$, $\phi^S_0$, $\xi_0$ and $\eta_0$, which are required  to build a non-trivial scalar potential
so as to realize the relevant  symmetry breaking.

\begin{table}[h]
	\begin{center}
		\begin{tabular}{|c||cccc|c|cccccc|c|cccc|}
			\hline
			\rule[14pt]{0pt}{0pt}
			& $l$ & $e^c$ & $\mu^c$ & $\tau^c$  & $h_{u,d}$ & $\phi _T $ & $\eta $ & $\tilde \eta $ & $\phi _S$ & $\xi $ & $\tilde \xi $ & $\Theta$ & $\phi ^T_0$ & $\eta_0$ & $\phi_0^S$ & $\xi_0$  \\ 
			\hline 
			\rule[12pt]{0pt}{0pt}
			$SU(2)$ & $2$ & $1$ & $1$ & $1$  & $2$ & $1$ & $1$ & $1$ & $1$ & $1$ & $1$ &$1$ & $1$ & $1$ & $1$ & $1$ \\
			$A_4$ & $\bf 3$ & $\bf 1$ & $\bf 1^{\prime\prime}$ & $\bf 1^\prime$ & $\bf 1$ & $\bf 3$ & $\bf 1''(1')$ & $\bf 1''(1')$ & $\bf 3$ & $\bf 1$ & $\bf 1$ &$1$ & $\bf 3$ & $\bf 1'(1'')$ & $\bf 3$ & $\bf 1$ \\
			$Z_3$ & $\omega $ & $\omega^2$ & $\omega^2$ & $\omega^2$  & $1$ & $1$ & $1$ & $1$ & $\omega$ & $\omega$ & $\omega$ &$1$ & $1$ & $\omega^2 $ & $\omega$ & $\omega$ \\
			$U(1)_{\rm FN}$ & $0$ & $4$ & $2$ &  $0$ & $0$ & $0$ & $0$ & $0$ & $0$ & $0$ &$0$ &$-1$ & $0$ & $0$ & $0$ & $0$ \\
			$U(1)_R$ & $1$ & $1$ & $1$ & $1$  & $0$ & $0$ & $0$ & $0$ & $0$ & $0$ & $0$ &$0$ & $2$ & $2$ & $2$ & $2$ \\
			\hline 
		\end{tabular}
	\end{center}
\vskip -0.3 cm
	\caption{Assignments of leptons, Higgs, flavons and driving fields,
	 where $\omega = \exp{(2\pi i/3)}$. }
	\label{tab:assignment}
\end{table}

In these setup, the superpotential for respecting $A_4 \times Z_3 \times 
U(1)_{\rm FN}\times U(1)_R$ symmetry 
is written by introducing the cutoff scale $\Lambda $  as 
\begin{align}
w&=w_Y+w_d, \nonumber \\
w_Y&=w_l+w_\nu , \nonumber \\
w_l&=y_e\left (\phi _Tl\right )_{\bf 1}e^ch_d\Theta ^4/\Lambda ^5+
y_\mu \left (\phi _Tl\right )_{\bf 1'}\mu ^ch_d\Theta ^2/\Lambda ^3
+y_\tau \left (\phi _Tl\right )_{\bf 1''}\tau^ch_d/\Lambda \nonumber \\
&+y'_e\left (\phi _Tl\right )_{\bf 1'(1'')}e^ch_d\eta \Theta ^4/\Lambda ^6+y'_\mu \left (\phi _Tl\right )_{\bf 1''(1)}\mu^ch_d\eta \Theta ^2/\Lambda ^4
+y'_\tau \left (\phi _Tl\right )_{\bf 1(1')}\tau^ch_d\eta /\Lambda ^2, \nonumber \\
w_\nu &=y_S(ll)_{\bf 3}h_uh_u\phi _S/\Lambda ^2+y_\xi(ll)_{\bf 1}h_uh_u\xi /\Lambda ^2 \nonumber \\
&+y'_1(ll)_{\bf 1}h_uh_u(\phi _S\phi _T)_{\bf 1}/\Lambda ^3+y'_2(ll)_{\bf 1'}h_uh_u(\phi _S\phi _T)_{\bf 1''}/\Lambda ^3 \nonumber \\
&+y'_3(ll)_{\bf 1''}h_uh_u(\phi _S\phi _T)_{\bf 1'}/\Lambda ^3+y'_4(ll)_{\bf 3}h_uh_u(\phi _S\phi _T)_{\bf 3}/\Lambda ^3 \nonumber \\
&+y'_5(ll)_{\bf 3}h_uh_u\phi _S \eta /\Lambda ^3+y'_6(ll)_{\bf 3}h_uh_u\xi \phi _T /\Lambda ^3+y'_7(ll)_{\bf 1'(1'')}h_uh_u\xi \eta /\Lambda ^3, \nonumber \\
w_d&=w_d^T+w_d^S, \nonumber \\
w_d^T&=-M\phi _0^T\phi _T+g\phi _0^T\phi _T\phi _T+\lambda \phi _0^T\phi _T\tilde \eta \nonumber \\
&-\lambda _1\eta _0\phi _T\phi _S+\lambda _2\eta _0\eta \xi +\lambda _3\eta _0\eta \tilde \xi 
+\lambda _4\eta _0\tilde \eta \xi +\lambda _5\eta _0\tilde \eta \tilde \xi , \nonumber \\
w_d^S&=g_1\phi _0^S\phi _S\phi _S +g_2\phi _0^S\phi _S\tilde \xi 
-g_3\xi _0\phi _S\phi _S+g_4\xi _0\xi \xi +g_5\xi _0\xi \tilde \xi +g_6\xi _0\tilde \xi \tilde \xi ,
\label{eq:superpotential}
\end{align}
where the subscripts $1'(1'')$ in $\left (\phi _Tl\right )_{\bf 1'(1'')}$, etc. correspond to 
the case of  $\eta $ for $1''(1')$.
The Yukawa couplings  $y$'s and $y'$'s are complex number of order one, 
 $M$ is  a complex mass parameter while 
$g$'s and $\lambda $'s are trilinear couplings which are also complex number of order one. 
Both leading operators and next-leading ones  are included in $w_Y$, 
which leads to the flavor structure of lepton mass matrices including next-leading corrections.
  
On the other hand, $w_d$ only contains leading operators,
 where we can force $\tilde \xi$($\tilde \eta$) to couple with
 $\phi _0^S\phi _S$ ($\phi _0^T\phi _T$),  but not  $\xi$($\eta$) with it
  since $\tilde \xi$ and $\xi$($\tilde \eta$ and $\eta$)
 have the same quantum numbers~\cite{Altarelli:2005yx}.
We can study the vacuum structure and lepton mass matrices
with these superpotential.
\subsection{Vacuum alignments of flavons}
Let us investigate the vacuum alignments of flavons.
The superpotentials $w_d^T$ and $w_d^S$ in Eq.~(\ref{eq:superpotential}) 
are written in terms of the components of triplet flavons:
\begin{align}
w_d^T&=-M\left (\phi _{01}^T\phi _{T1}+\phi _{02}^T\phi _{T3}+\phi _{03}^T\phi _{T2}\right )
+\lambda \left (\phi _{01}^T\phi _{T2}+\phi _{02}^T\phi _{T1}+\phi _{03}^T\phi _{T3}\right )\tilde \eta \nonumber \\
&+\frac{2g}{3}\left [\phi _{01}^T\left (\phi _{T1}^2-\phi _{T2}\phi _{T3}\right )+\phi _{02}^T\left (\phi _{T2}^2-\phi _{T1}\phi _{T3}\right )+\phi _{03}^T\left (\phi _{T3}^2-\phi _{T1}\phi _{T2}\right )\right ] \nonumber \\
&-\lambda _1\eta _0\left (\phi _{T2}\phi _{S2}+\phi _{T1}\phi _{S3}+\phi _{T3}\phi _{S1}\right )
+\lambda _2\eta _0\eta \xi +\lambda _3\eta _0\eta \tilde \xi +\lambda _4\eta _0\tilde \eta \xi 
+\lambda _5\eta _0\tilde \eta \tilde \xi , \nonumber \\
w_d^S&=\frac{2g_1}{3}\left [\phi _{01}^S\left (\phi _{S1}^2-\phi _{S2}\phi _{S3}\right )+\phi _{02}^S\left (\phi _{S2}^2-\phi _{S1}\phi _{S3}\right )
+\phi _{03}^S\left (\phi _{S3}^2-\phi _{S1}\phi _{S2}\right )\right ] \nonumber \\
&+g_2\left (\phi _{01}^S\phi _{S1}+\phi _{02}^S\phi _{S3}+\phi _{03}^S\phi _{S2}\right )\tilde \xi 
\nonumber \\
&-g_3\xi _0\left (\phi _{S1}^2+2\phi _{S2}\phi _{S3}\right )+g_4\xi _0\xi ^2+g_5\xi _0\xi \tilde \xi +g_6\xi _0\tilde \xi ^2,
\end{align}
where $w_d^S$ is the same superpotential given in Ref.~\cite{Altarelli:2005yx}.
Note that new terms including $\eta$ and $\tilde \eta$ are added in  $w_d^T$.

Then, the scalar potential of the $F$-term is given as
\begin{align}
V&\equiv V_T+V_S, \nonumber \\
V_T&=\sum _i\left |\frac{\partial w_d^T}{\partial \phi _{0i}^T}\right |^2+h.c. \nonumber \\
&=2\left |-M\phi _{T1}+\lambda \phi _{T2}\tilde \eta +\frac{2g}{3}\left (\phi _{T1}^2-\phi _{T2}\phi _{T3}\right )\right |^2 \nonumber \\
&+2\left |-M\phi _{T3}+\lambda \phi _{T1}\tilde \eta +\frac{2g}{3}\left (\phi _{T2}^2-\phi _{T1}\phi _{T3}\right )\right |^2 \nonumber \\
&+2\left |-M\phi _{T2}+\lambda \phi _{T3}\tilde \eta +\frac{2g}{3}\left (\phi _{T3}^2-\phi _{T1}\phi _{T2}\right )\right |^2 \nonumber \\
&+2\left |-\lambda _1\left (\phi _{T2}\phi _{S2}+\phi _{T1}\phi _{S3}+\phi _{T3}\phi _{S1}\right )
+\lambda _2\eta \xi +\lambda _3\eta \tilde \xi +\lambda _4\tilde \eta \xi 
+\lambda _5\tilde \eta \tilde \xi \right |^2  \ ,\nonumber \\
V_S&=\sum \left |\frac{\partial w_d^S}{\partial X}\right |^2+h.c. \nonumber \\
&=2\left |\frac{2g_1}{3}\left (\phi _{S1}^2-\phi _{S2}\phi _{S3}\right )+g_2\phi _{S1}\tilde \xi \right |^2
+2\left |\frac{2g_1}{3}\left (\phi _{S2}^2-\phi _{S1}\phi _{S3}\right )+g_2\phi _{S3}\tilde \xi \right |^2 \nonumber \\
&+2\left |\frac{2g_1}{3}\left (\phi _{S3}^2-\phi _{S1}\phi _{S2}\right )+g_2\phi _{S2}\tilde \xi \right |^2 \nonumber \\
&+2\left |-g_3\left (\phi _{S1}^2+2\phi _{S2}\phi _{S3}\right )+g_4\xi ^2+g_5\xi \tilde \xi +g_6\tilde \xi ^2\right |^2 \ .
\label{eq:scalar-potential}
\end{align}
The vacuum alignments of $\phi _T$, $\phi _S$ and VEVs of $\eta $, $\tilde \eta $, $\xi $ and $\tilde \xi $ 
are derived from the condition of the potential minimum,
that is $V_T=0$ and $V_S=0$ in Eq.(\ref{eq:scalar-potential}) as
\begin{align}
\langle \phi _T\rangle =&v_T(1,0,0),\quad \langle \phi _S\rangle =v_S(1,1,1),\quad \langle \eta \rangle =q, 
\quad \langle \tilde \eta \rangle =0, \quad \langle \xi \rangle =u,\quad \langle \tilde \xi \rangle =0, \nonumber \\
\nonumber \\
&v_T=\frac{3M}{2g},\qquad v_S^2=\frac{g_4}{3g_3}u^2, 
\qquad q=\frac{\lambda _1v_Tv_S}{\lambda _2u}=\frac{\lambda _1}{\lambda _2}\sqrt{\frac{g_4}{3g_3}}v_T \ ,
\label{eq:alignment}
\end{align}
where the VEVs of  $\tilde \xi$ and $\tilde \eta$ are  taken to be zero
 by the linear transformation of  $\xi$ and  $\tilde \xi$ ($\eta$ and  $\tilde \eta$)
 without loss of generality.
The coefficients $\lambda_i$ and $g_i$ are of  order one  since
 these flavons have no FN charges. 
 Therefore, the VEVs of $\eta$ and $\xi$  are  of  same order 
  as $v_T$ and $v_S$, respectively.
  In our numerical analyses,
   $q/\Lambda$ is scanned around  $v_T/\Lambda$ which  is fixed by the tau-lepton mass. 
   
   On the other hand,  the FN flavon $\Theta$ is not contained in $w_d$
   due to the $U(1)_{\rm FN}$ invariance.
    The VEV of  $\Theta$ can be derived from the scalar potential of $D$-term
    by assuming gauged $U(1)_{\rm FN}$.  The Fayet-Iliopolos term leads to the
    non-vanishing VEV of $\Theta$ as discussed in Ref.~\cite{Altarelli:2008bg}.
    Thus, its VEV is determined independently of $v_T$, $v_S$, $u$ and $q$.

\subsection{Lepton Mass Matrices}

The explicit lepton mass matrices are derived from the superpotentials
$w_l$  and $w_\nu$ in Eq.\,(\ref{eq:superpotential}) 
by use of the multiplication rule of $A_4$ given in Appendix B.
Let us begin with writing  down
the charged lepton mass matrices by imposing the vacuum alignments in Eq.(\ref{eq:alignment}) as:
\begin{equation}
M_\ell= v_d \alpha_\ell
\begin{pmatrix}
y_e \lambda^4 & 0 & y'_\tau \alpha_{\eta}\\
y'_e \alpha_{\eta}\lambda^4 & y_\mu\lambda^2 &0 \\0& y'_\mu \alpha_{\eta}\lambda^2& y_\tau
\end{pmatrix} \  {\rm for \ \eta(1'')}  \ , \quad 
v_d \alpha_\ell
\begin{pmatrix}
y_e \lambda^4 & y'_\mu \alpha_{\eta} \lambda^2  & 0\\
0 & y_\mu\lambda^2 &y'_\tau\alpha_{\eta} \\y'_e \alpha_{\eta}\lambda^4&0 & y_\tau
\end{pmatrix}   \ {\rm for \ \eta(1')}  \ ,
\end{equation}
where $\alpha_\ell$, $\alpha_\eta$ and $\lambda$ are defined in terms of the VEVs of  $\phi_T$, $\eta$ and $\Theta$, respectively:
\begin{equation}
\alpha_\ell\equiv \frac{\langle \phi_T \rangle}{\Lambda}=\frac{v_T}{\Lambda}\ ,
 \qquad 
 \alpha_{\eta}\equiv \frac{\langle \eta \rangle}{\Lambda}=\frac{q}{\Lambda}\ ,
 \qquad \lambda\equiv \frac{\langle \Theta \rangle}{\Lambda} \ .
\end{equation}
We note that  the off-diagonal elements arise from  the next-leading operators.

The left-handed mixing matrix of the charged lepton is derived 
by diagonalizing $M_\ell M_\ell^\dagger$.
We obtain the mixing matrix $U_\ell^\dagger$ approximately for the cases of $\eta$ being $1''$ or $1'$ of $A_4$ as
(more explicitly presented in Appendix C):
\begin{eqnarray}
U_\ell^\dagger &\simeq& \frac{1}{\sqrt{1+\alpha_\eta^{\tau^ 2}}}
\begin{pmatrix}
1 &  -{\cal O}(\alpha_{\eta}^2) & \alpha_{\eta}^\tau e^{i\varphi}\\
{\cal O}(\alpha_{\eta}^2) & 
\sqrt{1+\alpha_\eta^{\tau^ 2}}& {\cal O}(\alpha_{\eta} \lambda^4) \\
-  {\alpha_{\eta}^\tau} e^{-i\varphi}&  {\cal O}(\alpha_{\eta}^3) & 1
\end{pmatrix} \quad {\rm for \ \eta(1'')} \ ,  \nonumber \\
 \nonumber \\  \nonumber \\
U_\ell^\dagger &\simeq&  
\frac{1}{\sqrt{1+\alpha_\eta^{\mu 2}}} 
\frac{1}{\sqrt{1+\alpha_\eta^{\tau^ 2}}}
\begin{pmatrix}
\sqrt{1+\alpha_\eta^{\tau^ 2}} & \sqrt{1+\alpha_\eta^{\tau^ 2}}\alpha_{\eta}^\mu e^{i\varphi'}& {\cal O}(\alpha_{\eta}^2 \lambda^4)\\
- {\alpha_{\eta}^\mu}e^{-i\varphi'}& 1&\sqrt{1+\alpha_\eta^{\mu 2}} \alpha_{\eta}^\tau e^{i\varphi} \\
{\cal O}(\alpha_{\eta}^2)&  -{\alpha_{\eta}^\tau}e^{-i\varphi} & \sqrt{1+\alpha_\eta^{\mu 2}}
\end{pmatrix}  \quad  {\rm for \ \eta(1')},
\label{Uell}
\end{eqnarray}
where
\begin{equation}
 \alpha_{\eta}^\tau e^{i\varphi} \equiv \frac{y'_\tau}{y_\tau} \alpha_{\eta} \ , \qquad
 \alpha_{\eta}^\mu e^{i{\varphi}'}\equiv \frac{y'_\mu}{y_\mu} \alpha_{\eta} \ .
 \end{equation}
The mass eigenvalues $m_e^2$,  $m_\mu^2$ and $m_\tau^2$
are obtained by  $U_\ell M_\ell M_\ell^\dagger U_\ell^\dagger$ as shown in Appendix C.

In the leading order approximation, $U_\ell$ depends on one real parameter
$\alpha^\tau_\eta$  and one phase  $\varphi$ for the case of $\eta(1'')$, whereas it depends on
$\alpha^\tau_\eta$,  $\alpha^\mu_\eta$,    $\varphi$ and   $\varphi'$ 
for the case of $\eta(1')$.
The parameter $\alpha_{\eta}$ is expected to be much less than $1$ as discussed in the next section.
As seen in Eq.(\ref{Uell}), the off-diagonal (1,3)  and  (3,1) entries in
 $U_\ell^\dagger$ are dominant for the case of $\eta(1'')$ 
while the off-diagonal (1,2) and (2,3) (also (2.1) and (3,2)) entries in $U_\ell^\dagger$ are dominant for the case of  $\eta(1')$.
Thus, it is expected that the assignments of $\eta(1'')$ and $\eta(1')$ 
give rise to different predictions of the mixing and the CP violation.
It is found that the effects of the next-leading terms of
${\cal O} (\alpha_\eta^{2})$, ${\cal O} (\alpha_\eta^{3})$ and ${\cal O} (\alpha_\eta \lambda^4)$
in the mixing matrix $U_\ell^\dagger$ are negligibly small
by our numerical estimation.


The  neutrino mass matrix is derived from the superpotential $w_\nu$ 
in Eq.~(\ref{eq:superpotential}) 
by imposing the vacuum alignments given in Eq.(\ref{eq:alignment}).
The next-leading operator $y'_5 l l  h_u h_u\phi_S \eta$
can be absorbed in the leading one $y_S l l  h_u h_u\phi_S$  due to the alignment of $\langle \phi_S \rangle \propto (1,1,1)$.
Although the next-leading operators   $l l  h_u h_u\phi_S \phi_T$ and $l l  h_u h_u\phi_T \xi$ 
cannot be absorbed in the leading one,
their effects are expected to be suppressed
because $\langle \phi_T\rangle /\Lambda$ is fixed to be small.
 We have confirmed that the effect of those next-leading operators is negligibly small in our numerical calculations.
 
 On the other hand, the operator
$y'_7 l l  h_u h_u \xi\eta$ leads to
 the significant contribution to the neutrino mass matrix
because  $\langle\eta\rangle /\Lambda$ could be significantly larger than 
 $\langle \phi_T\rangle /\Lambda$ as discussed in Appendix D.
For $\eta(1'')$, we have
\begin{equation}
M_\nu=a
\begin{pmatrix}
1 & 0 &0 \\
0 & 1 &0 \\0& 0& 1
\end{pmatrix}
+ b
\begin{pmatrix}
1 & 1 &1 \\
1 & 1 &1 \\ 1& 1 & 1
\end{pmatrix}
+c
\begin{pmatrix}
1 & 0 &0 \\
0 & 0 &1 \\
0& 1& 0
\end{pmatrix}
+d
\begin{pmatrix}
0 & 1 &0 \\
1 & 0 & 0 \\ 0& 0& 1
\end{pmatrix} \ ,
\label{numassmatrix1}
\end{equation}
where the coefficients $a,b,c$ and $d$ are given in terms of the Yukawa couplings and VEVs of flavons as follows:
\begin{equation}
a=\frac{y_S\alpha_{\nu}}{\Lambda} v_u^2 \ , \quad
b=-\frac{y_S\alpha_{\nu}}{3\Lambda} v_u^2 \ , \quad
c=\frac{y_\xi\alpha_{\xi}}{\Lambda} v_u^2 \ , \quad 
d=\frac{y'_7\alpha_{\xi}\alpha_{\eta}}{\Lambda} v_u^2 \ , 
\label{abcd}
\end{equation}
with
\begin{equation}
\alpha_{\nu}\equiv \frac{\langle \phi_S \rangle}{\Lambda}
=\frac{v_S}{\Lambda}\ , \qquad
\alpha_{\xi}\equiv \frac{\langle \xi \rangle}{\Lambda}=\frac{u}{\Lambda}\ .
\label{alphaneu}
\end{equation}
Since the parameter $d$ is induced from the  next-leading operator $ll\xi\eta h_uh_u$,
 the magnitude of $d$ is expected to be much smaller than $a$, $b$ and $c$.
 
For $\eta(1')$,
 we get 
 \begin{equation}
 M_\nu=a
 \begin{pmatrix}
 1 & 0 &0 \\
 0 & 1 &0 \\0& 0& 1
 \end{pmatrix}
 + b
 \begin{pmatrix}
 1 & 1 &1 \\
 1 & 1 &1 \\ 1& 1 & 1
 \end{pmatrix}
 +c
 \begin{pmatrix}
 1 & 0 &0 \\
 0 & 0 &1 \\
 0& 1& 0
 \end{pmatrix}
 +d
 \begin{pmatrix}
 0 & 0 &1 \\
 0 & 1 & 0 \\ 1& 0& 0
 \end{pmatrix} \ ,
 \end{equation}
 where the last matrix of the right-hand side is a different one compared with
 the case of $\eta(1'')$.

There are three complex parameters in the model since the coefficient $b$ is given in terms of $a$.
We take $a$ to be real  without loss of generality and reparametrize them as follows:
\begin{equation}
a\rightarrow a \ , \quad  c\rightarrow c \ e^{i\phi_c} \ , \quad  
d\rightarrow d\  e^{i\phi_d} \ ,
\end{equation}
where $a$, $c$ and $d$ are real parameters and $\phi_c$, $\phi_d$ are CP violating phases.

For the lepton mixing matrix, Harrison-Perkins-Scott proposed a simple form of the mixing matrix,
so-called TBM mixing~\cite{Harrison:2002er,Harrison:2002kp},
\begin{equation}
V_{\text{TBM}}=
\begin{pmatrix}
\frac{2}{\sqrt{6}} & \frac{1}{\sqrt{3}} & 0 \\
-\frac{1}{\sqrt{6}} & \frac{1}{\sqrt{3}} & -\frac{1}{\sqrt{2}} \\
-\frac{1}{\sqrt{6}} & \frac{1}{\sqrt{3}} & \frac{1}{\sqrt{2}}
\end{pmatrix},
\label{UTBM}
\end{equation}
by which  $M_{\nu}$ is diagonalized  in the case of $d=0$.
We obtain the neutrino mass matrix in the TBM basis by rotating it with $V_{\rm TBM}$ as:
\begin{equation}
\hat M_\nu = V^T_{TBM} M_\nu  V_{TBM}
=
\begin{pmatrix}
a+c e^{i\phi_c}-\frac{d}{2} e^{i\phi_d} & 0 & \mp\frac{\sqrt{3}}{2}d e^{i\phi_d}\\
0 & ce^{i\phi_c}+de^{i\phi_d} &0 \\\mp\frac{\sqrt{3}}{2}d e^{i\phi_d}& 0& a-ce^{i\phi_c}+\frac{d}{2}e^{i\phi_d}
\end{pmatrix},
\label{neutrinomassmatrix}
\end{equation}
where upper (lower)  sign in front of (1,3) and (3,1) components correspond to  $\eta$ transforming as $1'' (1')$.
The neutrino mass eigenvalues are explicitly given in Appendix C.

The mixing matrix $U_\nu$ is derived from the diagonalization of $\hat M_\nu   \hat M_\nu^\dagger$ 
 apart from the Majorana phases such as
\begin{equation}
U_\nu \  (\hat M_\nu   \hat M_\nu^\dagger) \ U_\nu^\dagger = 
\begin{pmatrix}
m_1^2 & 0 & 0 \\
0& m_2^2 &0 \\0 & 0& m_3^2
\end{pmatrix}.
\end{equation}
As shown in Appendix C, we get 
\begin{equation}
U_\nu^\dagger = 
\begin{pmatrix}
\cos\theta & 0 & \sin\theta e^{-i\sigma}\\
0& 1&0 \\-\sin\theta e^{i\sigma} & 0& \cos\theta
\end{pmatrix},
\end{equation}
where  $\theta$ and $\sigma$
are given in terms of parameters in the neutrino  mass matrix.

As seen in Eq.(\ref{abcd}), the parameter $d$ is  related with $c$ as
\begin{equation}
\frac{d}{c}=\left |\frac{y'_7}{y_\xi} \right |\alpha_\eta  \equiv \alpha^\nu_\eta  \ ,
\end{equation}
where $y'_7$ and $y_\xi$ are coefficients of order one.
On the other hand, $a$ and $c$ are given
 in terms of $m_1$, $\alpha^\nu_\eta$ and the experimental data
$\Delta m^2_{\rm sol}$ and $\Delta m^2_{\rm atm}$ as shown in Appendix C.
Therefore,  $m_1$ and $\alpha^\nu_\eta$ are free parameters
as well as $\phi_c$ and $\phi_d$ in our model.


It is remarkable that
neutrino mass eigenvalues do not satisfy $\Delta m^2_{\rm sol}>0$ for the case of  IH of neutrino masses as discussed in Appendix C
because of the relation, $a\sim c$ and $c\gg d$, in our model.
It is understandable by considering the case of $d=0$ limit 
which corresponds to the exact TBM mixing.
It is allowed only for  NH of neutrino mass spectrum.


Finally, the Pontecorvo-Maki-Nakagawa-Sakata (PMNS) matrix~\cite{Maki:1962mu,Pontecorvo:1967fh}
is given as 
\begin{equation}
U_{\rm PMNS}=U_\ell\  V_{\rm TBM}\  U_\nu^\dagger\  P \ ,
\label{PMNSmatrix}
\end{equation}
where $P$ is the diagonal matrix responsible for the Majorana phases obtained from
\begin{equation}
P U_\nu \hat M_\nu U_\nu^TP={\rm diag}\{m_1, m_2, m_3\},
\end{equation}
where $m_1$, $m_2$ and $m_3$ are real positive neutrino masses.

The effective mass for the neutrinoless double beta ($0\nu\beta\beta$) decay is given as follows:
\begin{align}
|m_{ee}|=\left|m_1U^2_{e1}+m_2U^2_{e2}+m_3U^2_{e3}\right|  \ ,
\end{align}
where $U_{ei}$ denotes each component of the PMNS matrix $U_{\rm PMNS}$,
which includes the Majorana phases.

From Eq.(\ref{PMNSmatrix}), we can write down the three neutrino mixing angles of Appendix A
 in terms of our model parameters for the case of $1''$ singlet $\eta$, 
which shows how experimental results can be accommodated in our model:
 \begin{eqnarray}
&& \sin\theta_{12} \simeq \frac{1}{\sqrt{1+\alpha_\eta^{\tau^ 2}}}\frac{1}{\sqrt{3}} \left |1-\alpha_\eta^{\tau} e^{i\varphi}\right | \ ,  \nonumber\\
&&  \sin\theta_{13}\simeq \frac{1}{\sqrt{1+\alpha_\eta^{\tau^ 2}}} 
\left |\frac{2}{\sqrt{6}} \sin\theta e^{-i\sigma}-\frac{1}{\sqrt{2}}
 \alpha_\eta^{\tau}\cos\theta e^{i\varphi} \right |\ ,  \nonumber\\
 &&  \sin\theta_{23} \simeq \left |-\frac{1}{\sqrt{2}} \cos\theta
  -\frac{1}{\sqrt{6}} \sin\theta e^{-i\sigma}\right | \ ,  
  \label{Uappro}
 \end{eqnarray}
 where the next-leading terms are omitted.
 It is remarkable that  $\sin\theta_{13}$ is composed of contributions from both
the charged leptons and neutrinos.
 On the other hand,  the deviation from the trimaximal mixing of  $\theta_{12}$
 comes from the charged lepton sector, whereas  the deviation from the maximal mixing  of 
 $\theta_{23}$ comes from the neutrino sector.
 Since these are given in terms of four independent parameters,
   we cannot obtain the sum rules in the PMNS matrix elements.
   However,  the tau-lepton mass helps
   us to predict the allowed region of the CP violating Dirac phase $\delta_\text{CP}$
    and Majorana phases $\alpha_{21}$ and $\alpha_{31}$ as discussed in the next section.

\section{Numerical results}

\begin{table}[hbtp]
	\begin{center}
		\begin{tabular}{|c|c|c|}
			\hline 
			\rule[14pt]{0pt}{0pt}
			\  observable \ &  best fit and $1\,\sigma$  & $3\,\sigma$ range \\
			\hline 
			\rule[14pt]{0pt}{0pt}
			$\Delta m_{\rm atm}^2$& \ \   \ \ $(2.494^{+0.033}_{-0.031}) \times 10^{-3}{\rm eV}^2$ \ \ \ \
			&\ \ $(2.399\sim 2.593) \times 10^{-3}{\rm eV}^2$ \ \  \\
			\hline 
			\rule[14pt]{0pt}{0pt}
			$\Delta m_{\rm sol }^2$&   $(7.40^{+0.21}_{-0.20}) \times 10^{-5}{\rm eV}^2$
			& $(6.80\sim 8.02)  \times 10^{-5}{\rm eV}^2$ \\
			\hline 
			\rule[14pt]{0pt}{0pt}
			$\sin^2\theta_{23}$&  $0.538^{+0.033}_{-0.069}$ & $0.418\sim 0.613$ \\
			\hline 
			\rule[14pt]{0pt}{0pt}
			$\sin^2\theta_{12}$& $0.307^{+0.013}_{-0.012}$ & $0.272\sim 0.346$ \\
			\hline 
			\rule[14pt]{0pt}{0pt}
			$\sin^2\theta_{13}$&  $0.02206^{+0.00075}_{-0.00075}$ & $0.01981\sim 0.02436$ \\
			\hline 
		\end{tabular}
		\caption{The best fit,  $1\,\sigma$ and  $3\,\sigma$ ranges of neutrino oscillation parameters from NuFIT 3.2
			for NH \cite{NuFIT}. }
		\label{tab}
	\end{center}
\end{table}

At first, we present the framework of our calculations to predict the CP violating Dirac phase $\delta_\text{CP}$ and Majorana phases $\alpha_{21}$ and $\alpha_{31}$.
We explain how to get our predictions in terms of three real parameters
$\alpha^\tau_\eta$, $\alpha^\nu_\eta$ and $m_1$ on top of three phases $\varphi$, $\phi_c$ and $\phi_d$
for  NH of neutrino masses.
We can put for simplicity
\begin{equation}
\alpha_\eta=\alpha^\tau_\eta=\alpha^\nu_\eta \ ,
\end{equation}
that is $|y'_7/y_\xi|=|y'_\tau/y_\tau|=1$
since all Yukawa couplings are of order one.

The result of NuFIT 3.2~\cite{NuFIT} is used as input data to constrain the unknown parameters.
By taking $m_3^2-m_1^2=\Delta m_{\rm atm}^2$ and
 $ m_2^2-m_1^2=\Delta m_{\rm sol}^2$ with $3\,\sigma$ and  $1\,\sigma$ data given in Table 2,
$a$, $c$ and $d$ are fixed in terms of $m_1$, $\alpha_\eta$, $\phi_c$ and $\phi_d$. 
There is also the CP violating phase $\varphi$ in the charged lepton mixing matrix.
In our numerical analysis, we perform a parameter scan over those three phases and $m_1$ 
by generating random numbers. 
The scan ranges of the parameters are $-\pi \lesssim (\varphi, \phi_c, \phi_d)\lesssim \pi$ and $0\lesssim m_1 \lesssim 50 \mbox{~meV}$. Note that the range of $m_1$ is 
restricted by the lower bound of cosmological data for  the sum of neutrino masses,
 $160$~meV~\cite{Giusarma:2016phn}.
The parameter $\alpha_\eta$ is constrained by the tau-lepton mass:
\begin{equation}
m_\tau=|y_\tau| \alpha_\ell v_d \ ,
\label{alpha}
\end{equation}
which gives $\alpha_\ell=0.0316$ and $0.010$ for the minimal supersymmetric standard model (MSSM)  with $\tan\beta=3$ and SM, respectively. Here we put $|y_\tau|=1$.
 Since $\alpha_\eta$ is of same order  as $\alpha_\ell$ as seen in  Eq.(\ref{eq:alignment}),
 we vary the parameter  $\alpha_\eta$ around $\alpha_\ell=0.0316 \ (0.010)$  
 by using the  $\Gamma$ distribution ($\chi^2$ distribution),
 which is presented in Appendix D.

We calculate three neutrino mixing angles in terms of the model parameters
while keeping the parameter sets leading to values allowed by the experimental data at $1\,\sigma$ and $3\,\sigma$ C.L.
as given in Table 2.
Then, we calculate the CP violating phases and $|m_{ee}|$ with those selected parameter sets.
Accumulating enough parameter sets survived the above procedure, we make various scatter plots to show how observables depend on the model parameters.

In subsection 3.1, we show our numerical results for $\eta(1'')$.
The numerical results for $\eta(1')$ are briefly shown in subsection 3.2.

\subsection{Case of  a  $1''$ singlet  $\eta$}

\begin{figure}[h!]
	\begin{minipage}[]{0.47\linewidth}
		\includegraphics[{width=\linewidth}]{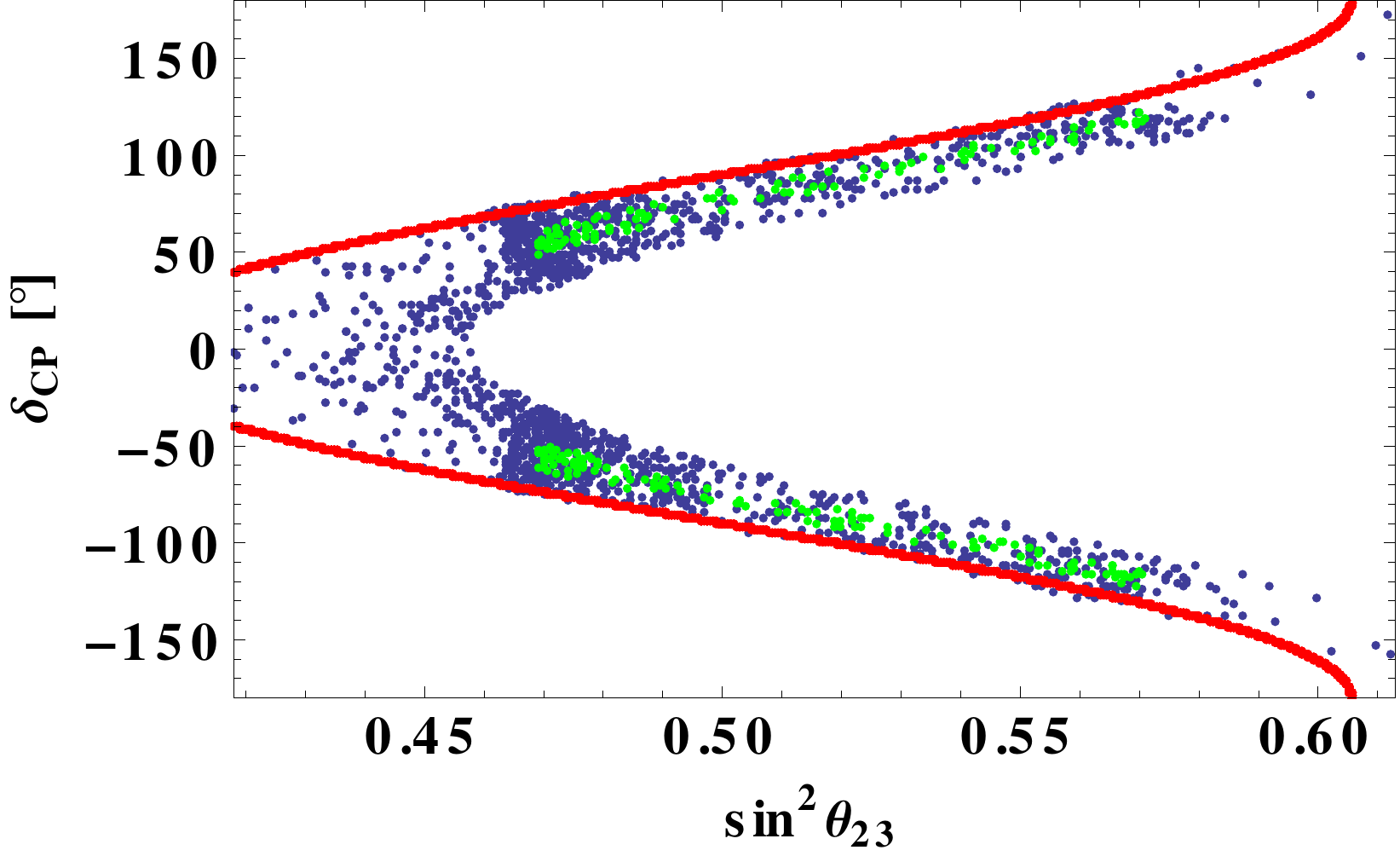}
		\caption{The allowed region on $\sin \theta _{23}$--$\delta _\text{CP}$ plane,
		where the blue and green dots correspond to the input of
		$3\,\sigma$ and  $1\,\sigma$ data  in Table 2,  respectively.
	The red curve represents the prediction of $\rm TM_2$.}
	\end{minipage}
	\hspace{5mm}
	\begin{minipage}[]{0.47\linewidth}
		\includegraphics[{width=\linewidth}]{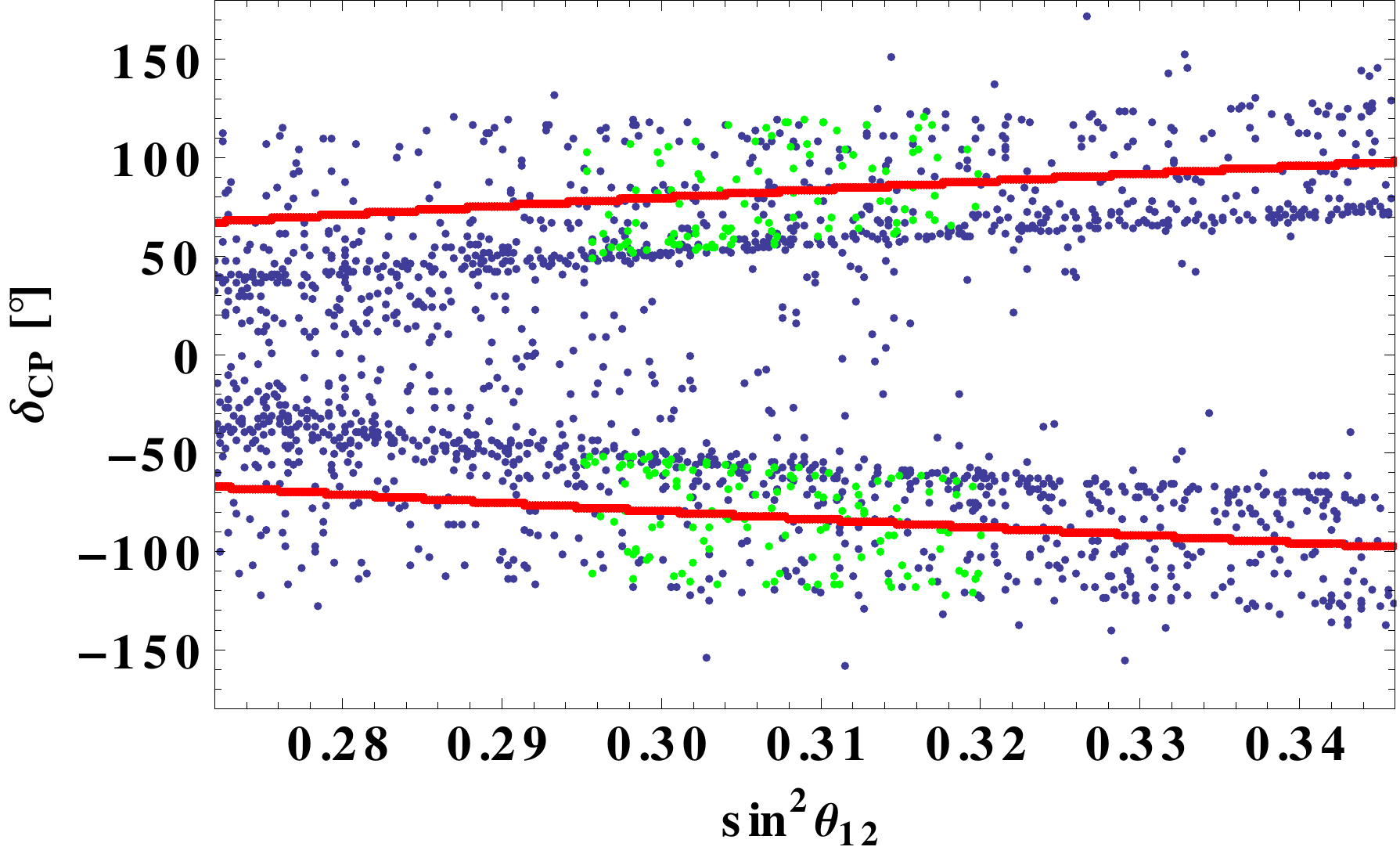}
		\caption{The allowed region on  $\sin ^2\theta _{12}$--$\delta _\text{CP}$ plane.
			The meaning of colors is the same as in Fig.1.
		The red curve represents the model without the  rotation to the neutrino mass matrix in the TBM basis.}
	\end{minipage}
\end{figure}

Let us show numerical results for the case of a $1''$ singlet  $\eta$.
We analyze only the case of  NH of neutrino masses
since the case of IH of neutrino masses is inconsistent
with experimental data as discussed in Appendix C.
 
At first, we show the prediction of $\delta_\text{CP}$ versus $\sin^2\theta_{23}$ 
 in Fig.~1 where the blue and green dots correspond to the input of
 $3\,\sigma$ and  $1\,\sigma$ data  in Table~2,  respectively.
 This result is similar to the prediction of ${\rm TM_2}$ 
  since the deviation from the  maximal mixing of $\theta_{23}$  is due to the extra (1-3) family rotation 
  of the neutrino mass matrix in Eq.(\ref{neutrinomassmatrix}). 
 In order to compare our prediction with the  $\rm TM_2$ result
 \cite{Shimizu:2014ria,Kang:2015xfa}, we draw its prediction
  by a red curve which is obtained by taking the best fit data in Table~2.
 We see that our predicted region is inside of the $\rm TM_2$ boundary.
  For the maximal mixing $\theta_{23}=\pi/4$, 
  the absolute value of $\delta_\text{CP}$ is expected to be $60^\circ$--$90^\circ$.
   It  is also  predicted to be $90^\circ\lesssim |\delta_\text{CP}|\lesssim 110^\circ$
   at the best fit of  $\sin^2\theta_{23}=0.538$.
All values between  $-180^\circ$ and  $180^\circ$ are allowed for $\delta_\text{CP}$    
in the case of the input data at $3\,\sigma$ as seen in Fig.~1.
  However, for the input data at $1\,\sigma$, $|\delta_\text{CP}|$ is restricted to be $50^\circ$--$120^\circ$, which is completely consistent with the present data at $1 \,\sigma$,  
$-157^\circ\lesssim \delta_\text{CP}\lesssim -83^\circ$ apart from its sign.
    Thus, the precise data of $\theta_{23}$ and $\delta_\text{CP}$ would
    provide us with a crucial test of our prediction.
  
 Next, we show the prediction of $\delta_\text{CP}$ versus $\sin^2\theta_{12}$ in Fig.~2.
 The  deviation from the trimaximal mixing of $\theta_{12}$  is due to 
the (1-3) family rotation of the charged lepton sector as seen in Eq.(\ref{Uappro}).
 The model without the additional rotation to the neutrino mass matrix in the TBM basis
  presented a clear correlation between $\sin^2\theta_{12}$ and $\delta_\text{CP}$~\cite{Shimizu:2014ria,Kang:2015xfa}.
 We also draw its prediction
 by a red curve which is obtained by taking the best fit data in Table~2.
  Predicted points are scattered  around the red curve. 
 Our predicted region is broad
 for the  $3\,\sigma$ data of mixing angles. 
 However, $1\,\sigma$ data forces
 the predicted region  to be rather narrow.
 Then, $|\delta_\text{CP}|=60^\circ$--$120^\circ$ is predicted 
 at the best fit of $\sin^2\theta_{12}=0.307$, where
  the maximal CP violation  $|\delta_\text{CP}|=90^\circ$ is still allowed.
 
On the other hand, we cannot find any correlation between $\delta_\text{CP}$ and $\sin^2\theta_{13}$ 
since both phases $\sigma$ in the neutrino mass matrix and $\varphi$ 
in the charged lepton mass matrix contribute
 to $\sin^2\theta_{13}$ as seen in Eq.(\ref{Uappro}).
 We omit to present the result in a figure.

\begin{figure}[h!]
	\begin{minipage}[]{0.47\linewidth}
		\includegraphics[{width=\linewidth}]{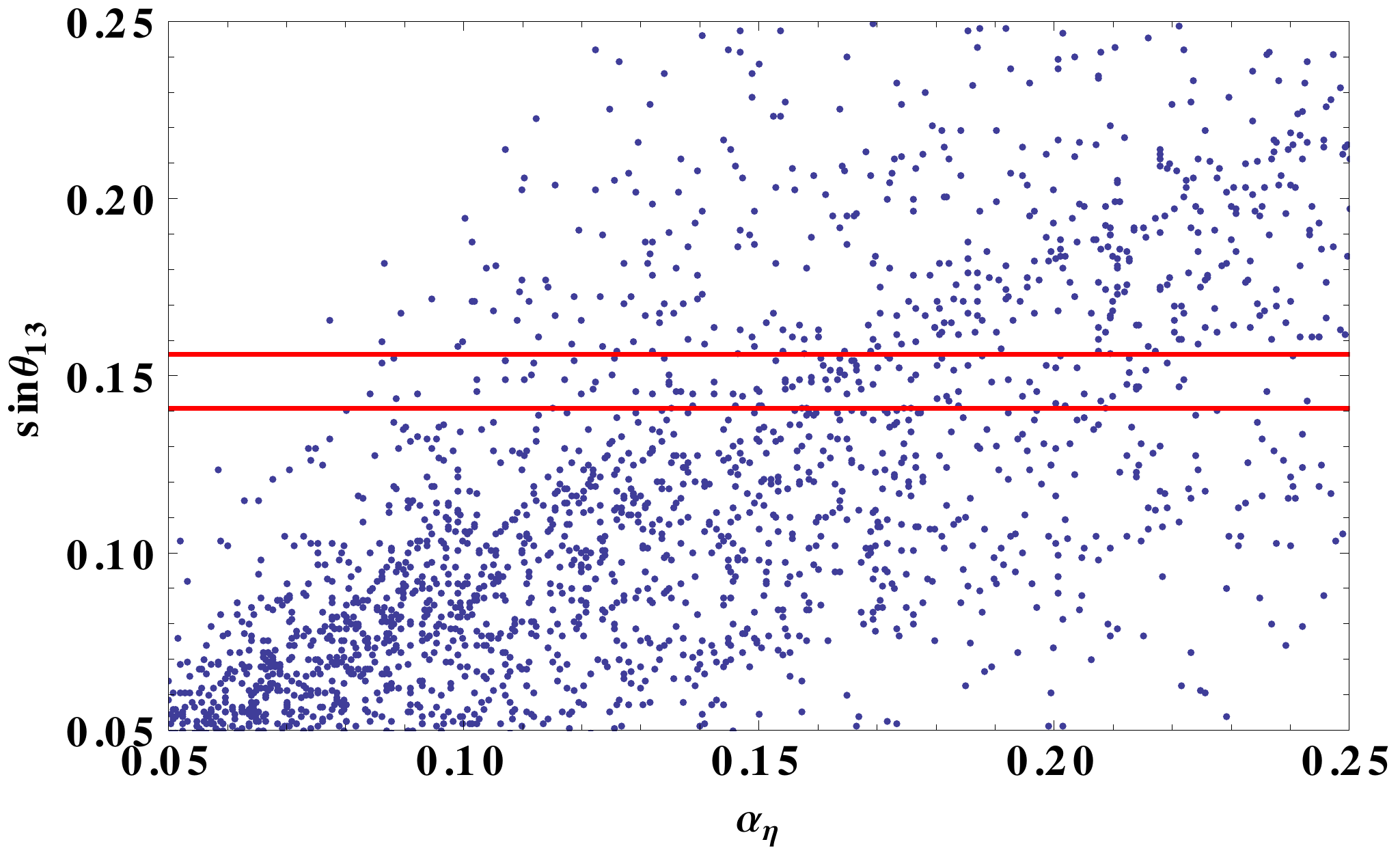}
		\caption{The allowed region on $\alpha _\eta $--$\sin \theta _{13}$ plane,
			where the $3\,\sigma$ data is taken except for $\sin \theta _{13}$.
			The red lines represent the upper and lower bounds of the experimental data.}	
	\end{minipage}
	\hspace{5mm}
	\begin{minipage}[]{0.47\linewidth}
		\includegraphics[{width=\linewidth}]{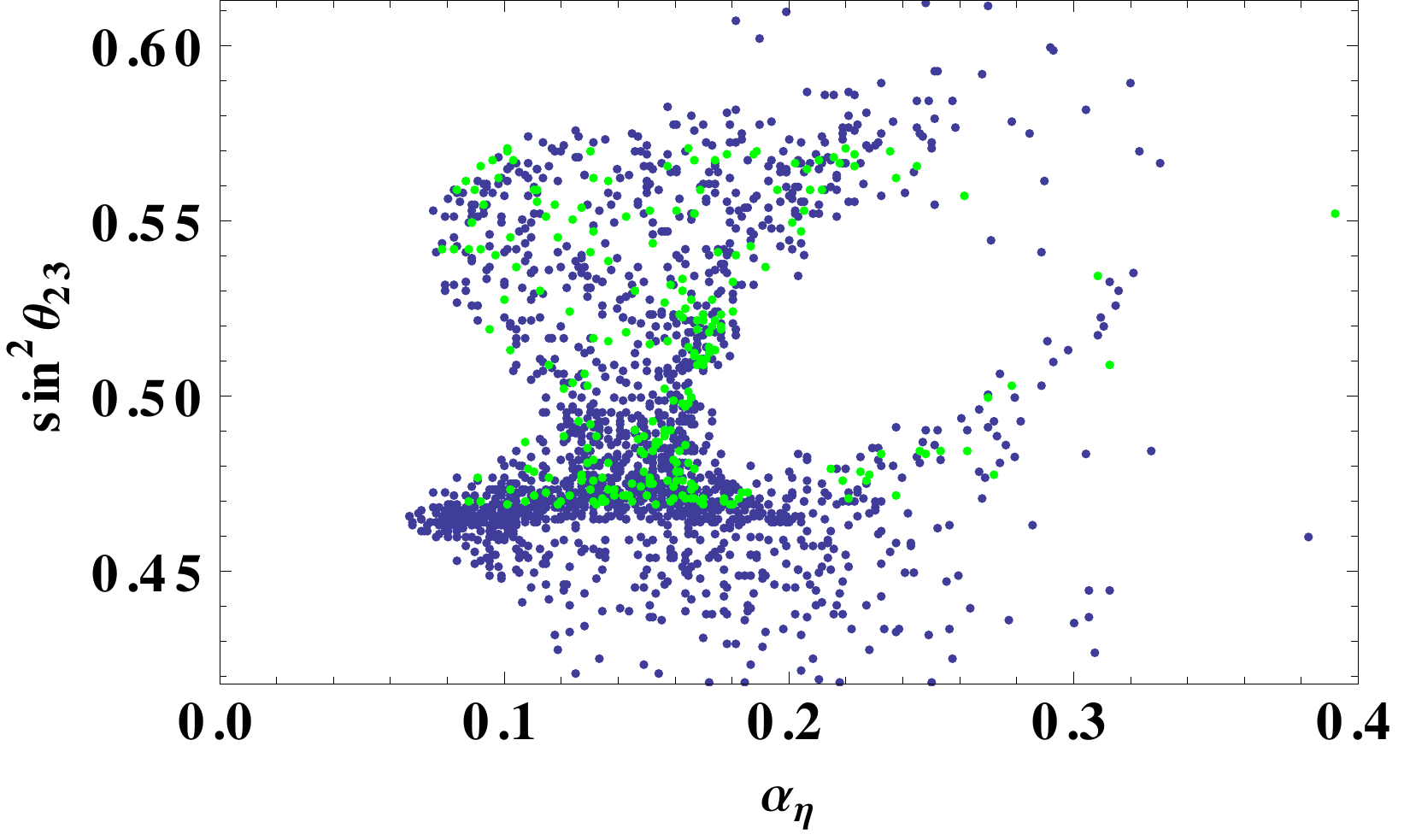}
		\caption{The allowed region on  $\alpha _\eta $--$\sin ^2\theta _{23}$ plane. 
			The meaning of colors is the same as in Fig.~1.}
	\end{minipage}
\end{figure}

\begin{figure}[h!]
	\begin{minipage}[]{0.47\linewidth}
		\includegraphics[{width=\linewidth}]{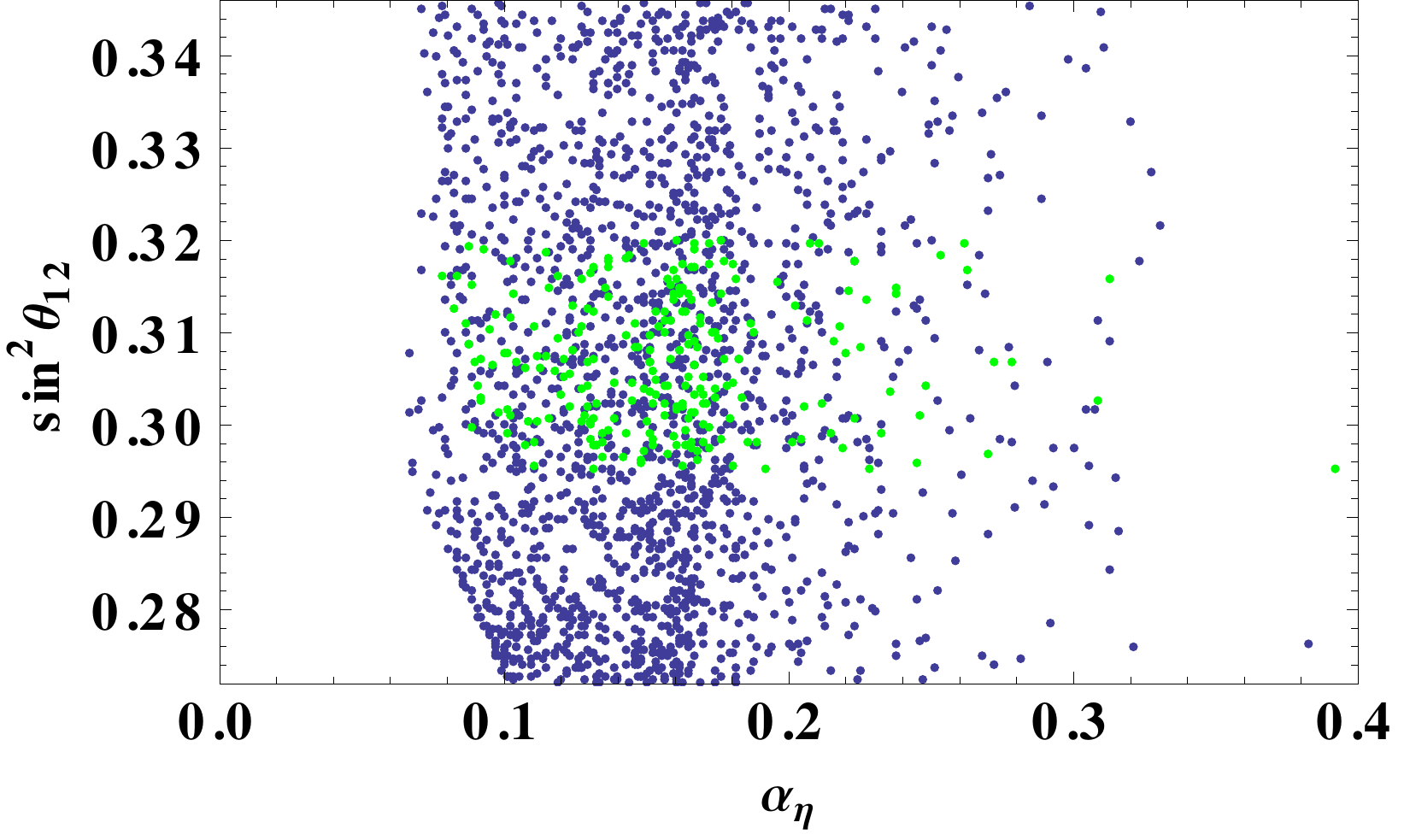}
		\caption{The allowed region on $\alpha _\eta$--$\sin ^2\theta _{12}$ plane.
		The meaning of colors is the same as in Fig.~1.	}
	\end{minipage}
	\hspace{5mm}
	\begin{minipage}[]{0.47\linewidth}
		\vspace{-5mm}
		\includegraphics[{width=\linewidth}]{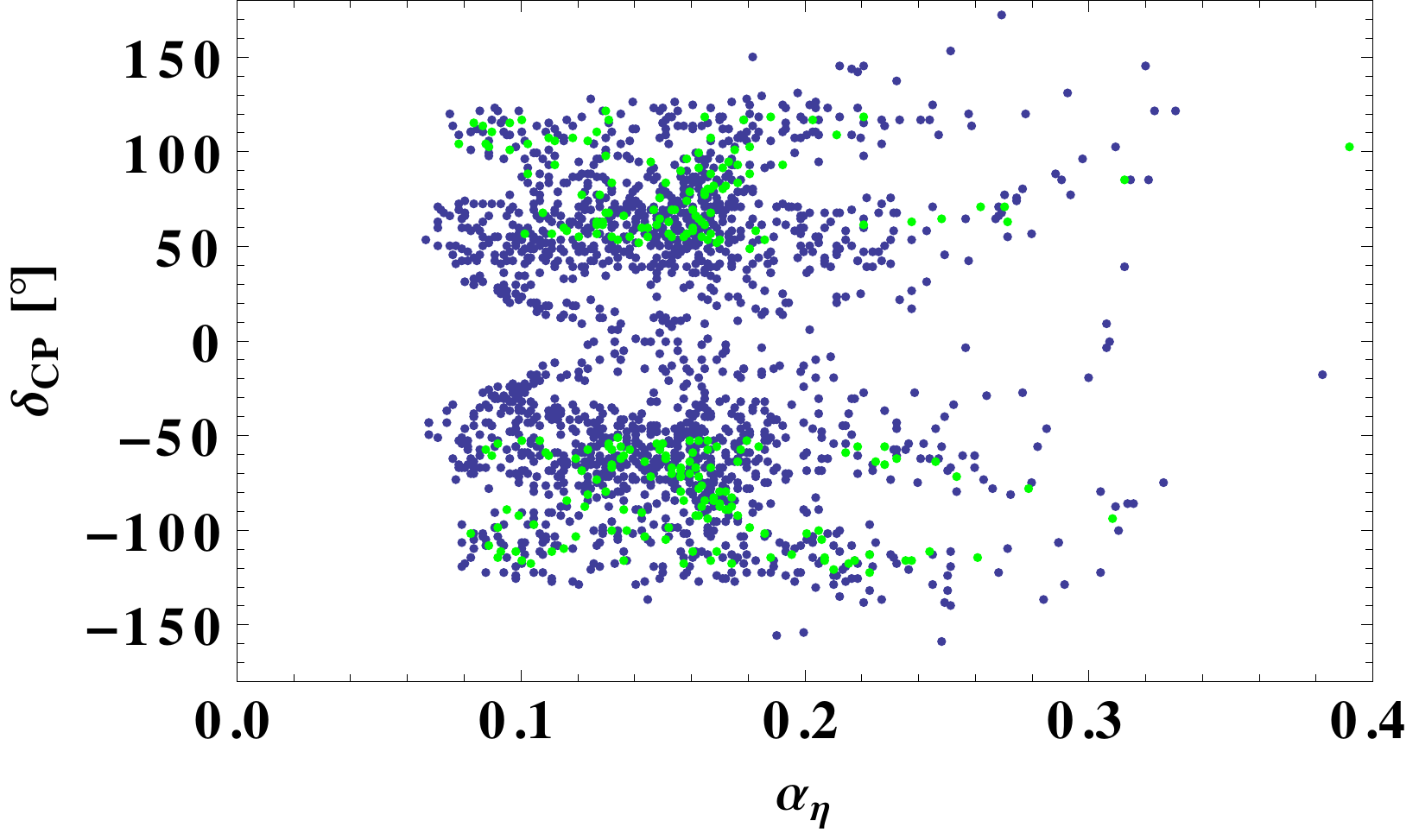}
		\caption{The allowed region on  $\alpha _\eta $--$\delta _\text{CP}$ plane.
                The meaning of colors is the same as in Fig.~1.}
	\end{minipage}
\end{figure}

In order to understand the role of the key parameter $\alpha_\eta$,
we show how  the three neutrino mixing angles and
 the CP violating Dirac phase depend on $\alpha_\eta$ in Figs.\,3--6.
 At first, in Fig.\,3,
  we show the prediction of $\sin\theta_{13}$ versus $\alpha_\eta$ 
where  the $3\,\sigma$ data is  taken as input  except for $\sin \theta _{13}$.
 The red lines denote the upper and lower bounds of the $3\,\sigma$ experimental data for $\sin \theta_{13}$.
 Note that $\sin\theta_{13}$ depends  on $\alpha_\eta$ crucially as seen in Eq.(\ref{Uappro}).
 As shown in Fig.\,3, the observed value $\sin\theta_{13}$ is not reproduced
 unless $\alpha_\eta$ is larger than $0.07$.

The clear dependence between $\alpha_\eta$ and the predicted $\sin^2\theta_{23}$
can be seen in Fig.4.
In order to reproduce  the  maximal mixing of $\theta_{23}$,
$\alpha_\eta$ should be larger than $0.12$.
The highly probable  prediction of $\sin^2\theta_{23}$ is
 near  $0.47-0.5$ for $0.1\leq \alpha_\eta \leq 0.2$.
 
      The deviation from the trimaximal mixing  of  $\sin^2\theta_{12}$ explicitly depends on $\alpha_\eta$ as seen in Eq.(\ref{Uappro}).
      We show  the prediction of $\sin^2\theta_{12}$ versus  $\alpha_\eta$ in Fig.~5.
      The predicted $\sin^2\theta_{12}$ is almost  independent of $\alpha_\eta$
      as far as $\alpha_\eta\geq 0.1$.

    The  $\alpha_\eta$ dependence on $\delta_\text{CP}$
    gives the characteristic prediction as shown  in Fig.~6. 
    The CP conservation $\delta_\text{CP}=0$ is excluded
     in the smaller region $\alpha_\eta\leq 0.12$ for 
      the experimental data with $3\,\sigma$.
     By inputting the  $1\,\sigma$ data in Table~2,
    we  obtain the prediction of  $\delta_\text{CP}$ to be $\pm (50^\circ - 120^\circ)$,
    which is almost independent of $\alpha_\eta$ for $\alpha_\eta=0.1$--$0.2$.

  \begin{figure}[t!]
  	\begin{minipage}[]{0.47\linewidth}
  		\includegraphics[{width=\linewidth}]{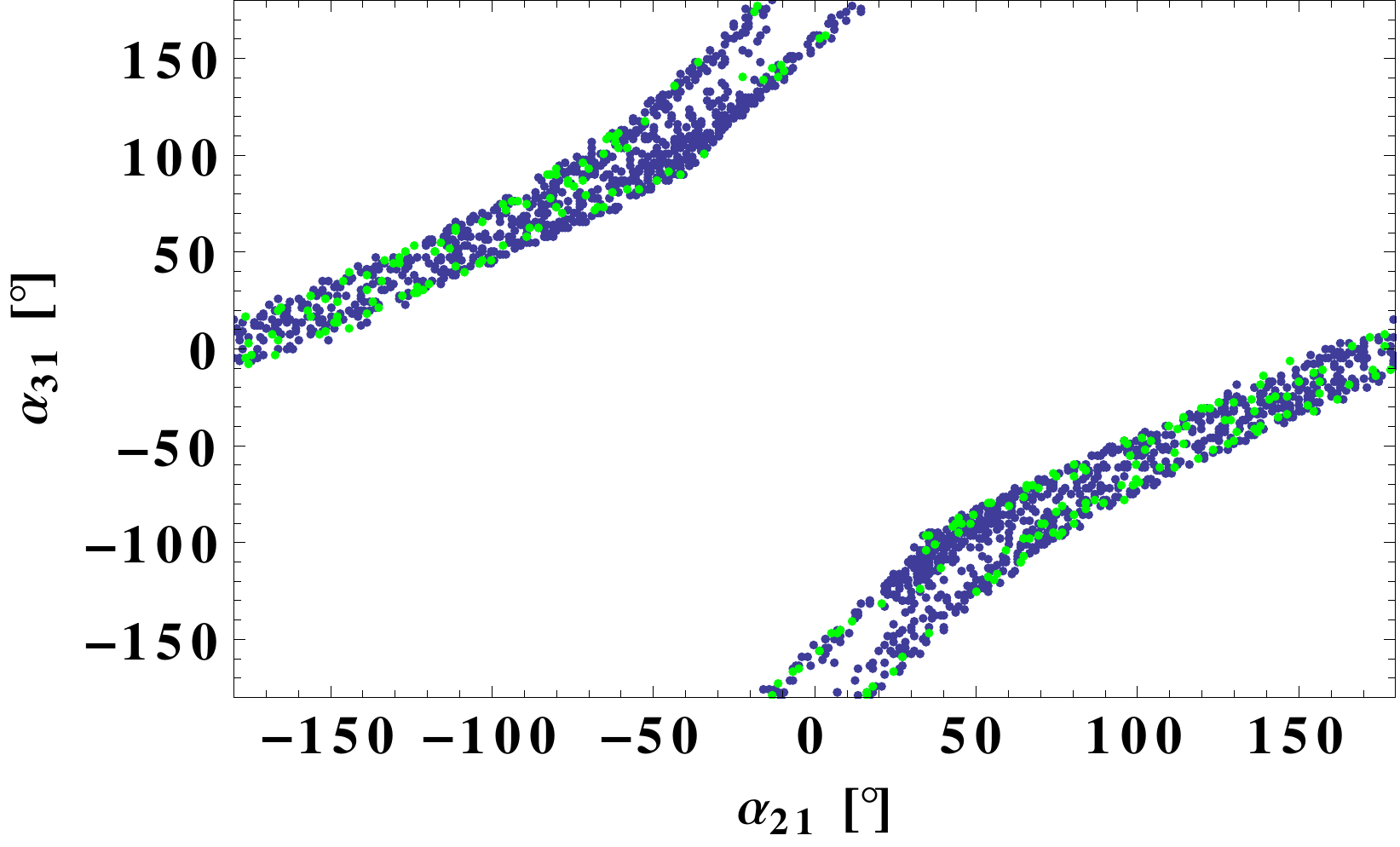}
  		\caption{The predicted Majorana phases on $\alpha _{21}$--$\alpha _{31}$ plane.
The meaning of colors is the same as in Fig.~1.}
  	\end{minipage}
  	\hspace{5mm}
  	\begin{minipage}[]{0.47\linewidth}
		\vspace{-7mm}
  		\includegraphics[{width=\linewidth}]{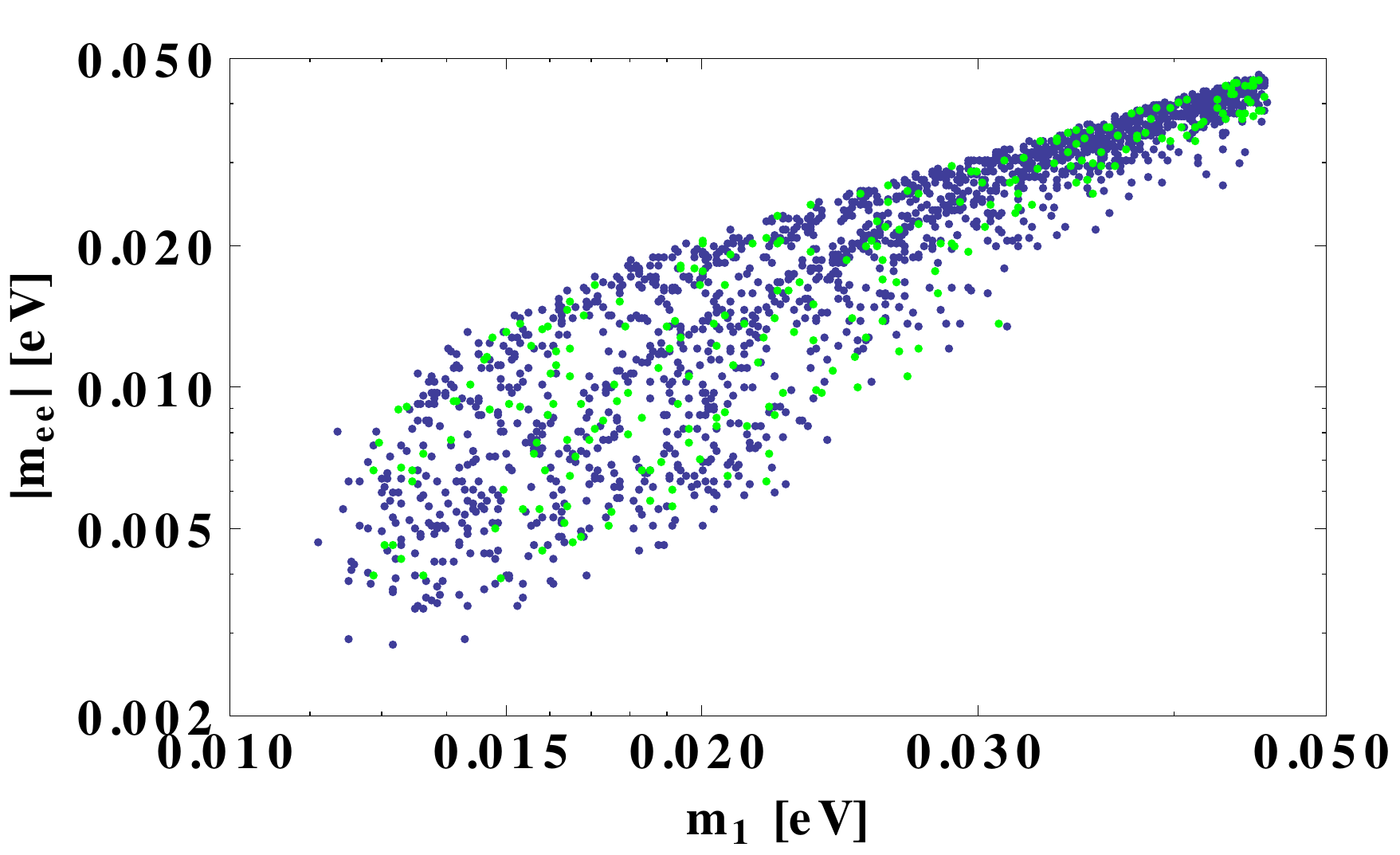}
  		\caption{The prediction of $|m_{ee}|$ vs. $m_1$.
  			The meaning of colors is the same as in Fig.~1.}
  	\end{minipage}
  \end{figure}
  \begin{figure}[t!]
  	\begin{minipage}[]{0.47\linewidth}
  		\includegraphics[{width=\linewidth}]{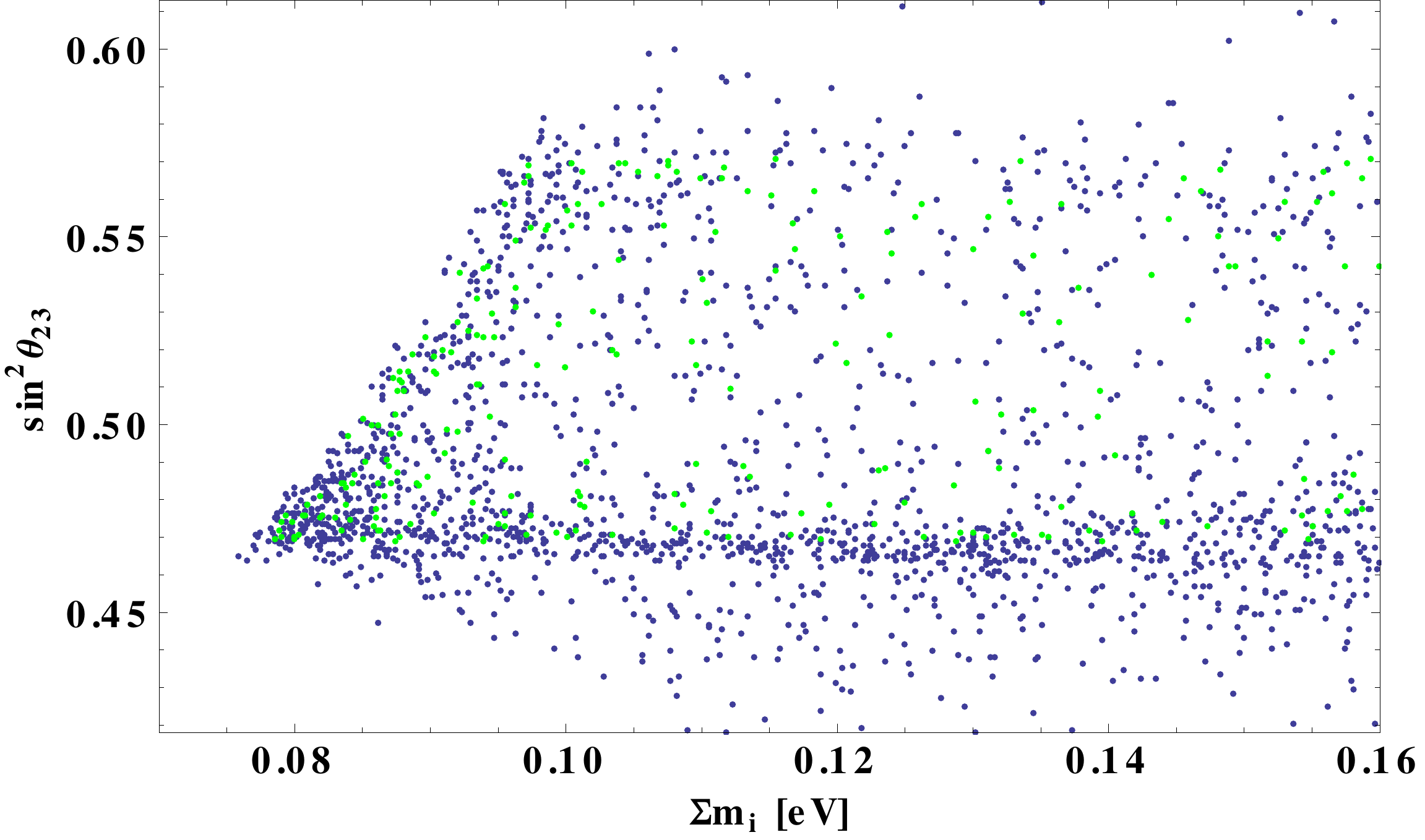}
  		\caption{The $\Sigma m_i$ dependence of the predicted  $\sin^2\theta_{23}$.
The meaning of colors is the same as in Fig.~1.}
  	\end{minipage}
  	\hspace{5mm}
  	\begin{minipage}[]{0.47\linewidth}
  		\includegraphics[{width=\linewidth}]{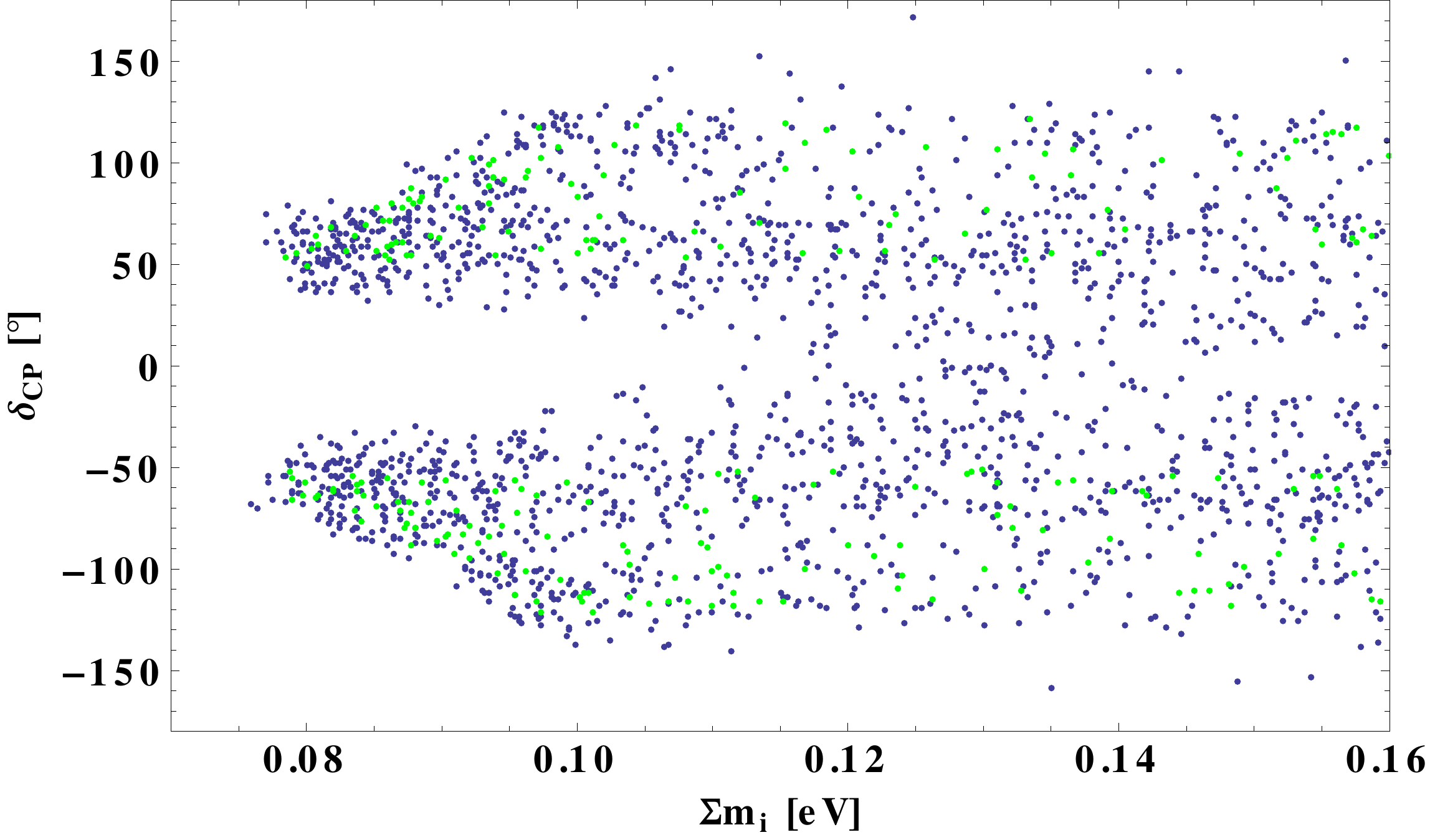}
  		\caption{The $\Sigma m_i$ dependence of the predicted $\delta_\text{CP}$.
The meaning of colors is the same as in Fig.~1.}
  	\end{minipage}
  \end{figure}
 We show the prediction of the Majorana phases $\alpha_{21}$ and $\alpha_{31}$
 in Fig.~7.
 While both Majorana phases are allowed in all region of $-180^\circ \sim 180^\circ$,
 there is a clear correlation between both phases.
 
In Fig.~8, we present the predicted $|m_{ee}|$, the effective mass for the $0\nu\beta\beta$ decay, versus $m_1$ which is another key parameter in our model.
  The parameter $m_1$ should be larger than $12$\,meV 
 in order to reproduce the observed mass squared differences,
 and it is smaller than $46$\,meV due to the cosmological constraint on the sum of neutrino masses\,\cite{Giusarma:2016phn}.
 In the hierarchical case of neutrino masses $m_1< m_2\ll m_3$,
 the predicted value $|m_{ee}|$ is at most $10$\,meV but close to  $45$\,meV
 for the degenerated  neutrino masses.

  Next, we discuss the sum of three neutrino masses $\Sigma m_i$
  because the cosmological observation gives us a upper bound  for it.
    We show the predicted region of $\Sigma m_i$ -- $\sin^2\theta_{23}$ plane
    in Fig.~9. 
    The minimum of the sum of three neutrino masses $\Sigma m_i$ is $75$\,meV in our model.
    In order to get $\sin^2\theta_{23}\geq 0.5$,
     $\Sigma m_i$ should be larger than $85$\,meV.
      For the best fit of  $\sin^2\theta_{23}=0.538$, 
     $\Sigma m_i$ is expected to  be larger than $90$\,meV.
    We show the predicted region of $\Sigma m_i$ -- $\delta_\text{CP}$ plane
    in Fig.~10.
    The predicted $|\delta_\text{CP}|$ is smaller than $90^\circ$
     if $\Sigma m_i$ is smaller than $85$\,meV.
     Thus, the cosmological observation for the sum of neutrino masses
     will be a crucial test of these predictions.

We have neglected the next-leading terms 
  $l l\phi_S \phi_T h_u h_u$ and $l l \phi_T \eta h_u h_u$ in the neutrino mass matrix
  of Eq.(\ref{numassmatrix1})
 because $\alpha_\ell=0.0316 \ (0.010)$ is  small
 compared with $\alpha_\eta\geq 0.1$. 
We have confirmed that those effects are  small with our numerical calculation
by inputting $1\,\sigma$ data.
Indeed, the prediction of  $\sin^2\theta_{23}$--$\delta_\text{CP}$
 almost remains inside of the red curve in Fig.~1.
 
 It is also deserved to comment on the $\alpha_\eta$ distribution  in our numerical results.
 In order to remove the predictions for $\alpha_\eta>0.3$ smoothly, 
 which is about ten times larger than  $\alpha_\ell= 0.0316$, we have used the Gamma distribution for $\alpha_\eta$  given in Eq.\,(\ref{Gamma1}) of Appendix D.
  We have confirmed   that our results are not  changed even if 
   we adopt another Gamma distribution presented  in Eq.(\ref{Gamma2})
  of Appendix D although the number density of dots gets lower.
 We have also used $\alpha_\ell=0.010$ which corresponds to SM in our calculations.
   In this case, the number density of dots significantly gets lower,
   but the allowed region is almost unchanged.
   Moreover, we have found that  the allowed region is also unchanged even if
    we use  the flat-distribution of  $\alpha_\eta$ in the region
    $0\leq \alpha_\eta \leq 0.3$.
    Thus, our results are robust 
    for any  distribution of $\alpha_\eta$.

 \subsection{Case of a $1'$ singlet $\eta$}
 
 \begin{figure}[t]
 	\begin{minipage}[]{0.47\linewidth}
 		\includegraphics[{width=\linewidth}]{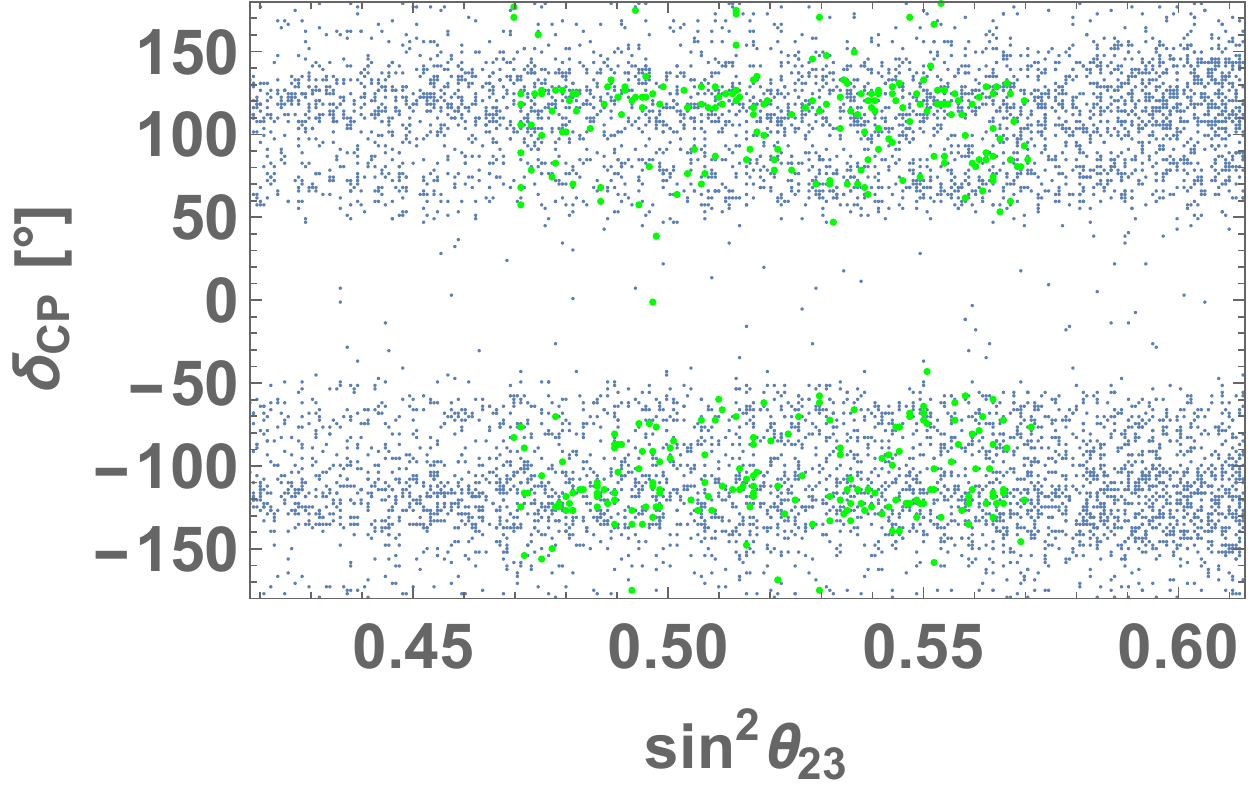}
 		\caption{The allowed region on $\sin \theta _{23}$--$\delta _\text{CP}$ plane
 			for $\eta(1')$. The meaning of colors is the same as in Fig.~1.}
 	\end{minipage}
 	\hspace{5mm}
 	\begin{minipage}[]{0.47\linewidth}
 		\includegraphics[{width=\linewidth}]{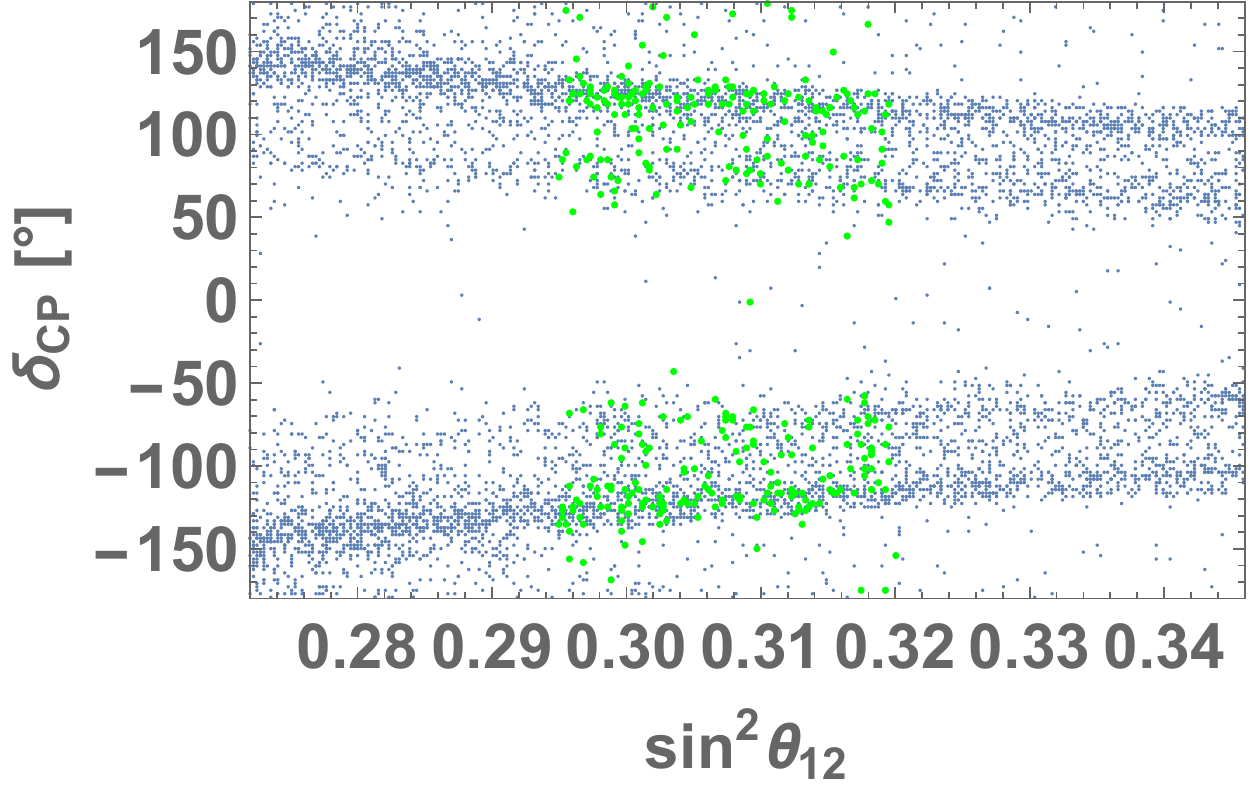}
 		\caption{The allowed region on  $\sin ^2\theta _{12}$--$\delta _\text{CP}$ plane
 				for $\eta(1')$. The meaning of colors is the same as in Fig.~1.}
 	\end{minipage}
 \end{figure}
 We show the numerical results  for  a $1'$ singlet $\eta$  briefly
 because the correlations of  the observables appear to be weak.
 We show the predicted $\delta_\text{CP}$ versus $\sin^2\theta_{23}$ 
 in Fig.\,11. 
 The region of $|\delta_\text{CP}|\leq 50^\circ$ is almost excluded 
 while the regions near  $\pm 180^\circ$ are allowed.
There is no  correlation between  $\sin^2\theta_{23}$ and $\delta_\text{CP}$.
 
 We also show the predicted  $\delta_\text{CP}$ versus $\sin^2\theta_{12}$ 
    in Fig.\,12.
   The predicted $|\delta_\text{CP}|$ increases
    as $\sin^2\theta_{12}$ decreases from the trimaximal mixing $1/3$,
    but its correlation is rather weak.

Both results in Figs.\,11 and 12
  are due to both mixing of (1-2) and (2-3) families 
 in  the charged lepton sector.
 Thus, the model  with the $1'$ singlet  $\eta$ is less attractive than that with the $1''$ singlet $\eta$ 
  in  light of NuFIT 3.2 data.

\section{Summary}

The flavor symmetry of leptons can be examined precisely in light of the new data and the upcoming experiments~\cite{Petcov:2018snn}.
We study the $A_4$ model with minimal parameters  by using the results of  NuFIT 3.2.
We introduce the $A_4$ singlet $1'$ or $1''$ flavon $\eta$
which couples to both the charged  lepton  and neutrino sectors in the next-leading order 
due to the relevant  $Z_3$ charge for $\eta$.
The model with the  $1''(1')$ flavon is
consistent with the experimental data of $\Delta m_{\rm sol }^2$
 only for NH of neutrino masses.
 The key parameter is $\alpha_\eta$ which is derived from the VEV of the flavon $\eta$.
  The parameter  $\alpha_\eta$ is distributed
  around $\alpha_\ell=0.0316 \ (0.010)$ in the  Gamma distribution of the statistic.
  Our results are robust for different distribution of $\alpha_\eta$.
  
In the case of  the singlet $\eta(1'')$, 
 $\alpha_\eta$ should be larger than $0.07$ in order to reproduce the observed value of $\sin\theta_{13}$. 
The numerical prediction of 
$\delta_\text{CP}$ versus $\sin^2\theta_{23}$ is similar to
the prediction of ${\rm TM_2}$.
However, 
our predicted region is inside of the $\rm TM_2$ boundary.
The absolute value of the predicted $\delta_\text{CP}$ is $60^\circ$--$90^\circ$
for the maximal mixing $\theta_{23}=\pi/4$.
For the best fit of  $\sin^2\theta_{23}=0.538$, $|\delta_\text{CP}|$ is 
in the region of $90^\circ$--$110^\circ$. 
The predicted  $\sin^2 \theta_{12}$ is also allowed  around
the  best fit of NuFIT 3.2
while  keeping near the maximal mixing of  $\theta_{23}$.
Inputting the data with the $1\,\sigma$ error-bar, we obtain the clear prediction
of the CP violating  Dirac phase to be $|\delta_\text{CP}|=50^\circ- 120^\circ$.
The lightest neutrino mass   $m_1$ is expected to be $12$\,meV--$46$\,meV,
which leads to the $|m_{ee}|<45$\,meV. 
 In order to get the best fit of  $\sin^2\theta_{23}=0.538$, 
the sum of the three neutrino masses is expected to  be larger than $90$\,meV.
The cosmological observation  for the  sum of the neutrino masses
will also provide  a crucial test of these predictions.

The model with $\eta(1')$ is not attractive
 in  light of NuFIT 3.2 result because the input data given in Table~2
 does not give a severe constraint for the predicted region of $\delta_\text{CP}$.
 
 We expect the precise measurement of the CP violating phase 
 to test the model in the future.

\vspace{0.5cm}
\noindent
{\bf Acknowledgement}  

This work is supported by JSPS Grants-in-Aid for Scientific Research
16J05332 (YS) and 15K05045, 16H00862 (MT).
S.K.Kang was supported by the National Research Foundation of Korea (NRF) grants (2009-0083526,
2017K1A3A7A09016430, 2017R1A2B4006338).


\appendix

\section*{Appendix}

\section{Lepton mixing matrix}

Supposing neutrinos to be Majorana particles, 
the PMNS matrix $U_{\text{PMNS}}$~\cite{Maki:1962mu,Pontecorvo:1967fh} 
is parametrized in terms of the three mixing angles $\theta _{ij}$ $(i,j=1,2,3;~i<j)$,
one CP violating Dirac phase $\delta _\text{CP}$ and two Majorana phases 
$\alpha_{21}$, $\alpha_{31}$  as follows:
\begin{align}
U_\text{PMNS} =
\begin{pmatrix}
c_{12} c_{13} & s_{12} c_{13} & s_{13}e^{-i\delta _\text{CP}} \\
-s_{12} c_{23} - c_{12} s_{23} s_{13}e^{i\delta _\text{CP}} &
c_{12} c_{23} - s_{12} s_{23} s_{13}e^{i\delta _\text{CP}} & s_{23} c_{13} \\
s_{12} s_{23} - c_{12} c_{23} s_{13}e^{i\delta _\text{CP}} &
-c_{12} s_{23} - s_{12} c_{23} s_{13}e^{i\delta _\text{CP}} & c_{23} c_{13}
\end{pmatrix}
\begin{pmatrix}
1&0 &0 \\
0 & e^{i\frac{\alpha_{21}}{2}} & 0 \\
0 & 0 & e^{i\frac{\alpha_{31}}{2}}
\end{pmatrix},
\label{UPMNS}
\end{align}
where $c_{ij}$ and $s_{ij}$ denote $\cos\theta_{ij}$ and $\sin\theta_{ij}$, respectively.

The rephasing invariant CP violating measure, Jarlskog invariant~\cite{Jarlskog:1985ht},
is defined by the PMNS matrix elements $U_{\alpha i}$. 
It is written in terms of the mixing angles and the CP violating phase as:
\begin{equation}
J_{CP}=\text{Im}\left [U_{e1}U_{\mu 2}U_{e2}^\ast U_{\mu 1}^\ast \right ]
=s_{23}c_{23}s_{12}c_{12}s_{13}c_{13}^2\sin \delta _\text{CP}~ ,
\label{Jcp}
\end{equation}
where $U_{\alpha i}$ denotes the each component of the PMNS matrix.

There are also other invariants $I_1$ and $I_2$ associated with Majorana phases
\cite{Bilenky:2001rz}-\cite{Girardi:2016zwz},
\begin{equation}
I_1=\text{Im}\left [U_{e1}^\ast U_{e2} \right ]
=c_{12}c_{12}c_{13}^2\sin \left (\frac{\alpha_{21}}{2}\right )~, \quad
I_2=\text{Im}\left [U_{e1}^\ast U_{e3} \right ]
=c_{12}s_{13}c_{13}\sin \left (\frac{\alpha_{31}}{2}-\delta_\text{CP}\right )~.
\label{Jcp}
\end{equation}
We calculate $\delta_\text{CP}$, $\alpha_{21}$ and $\alpha_{31}$ with these relations.

\section{Multiplication rule of $A_4$ group}
\label{sec:multiplication-rule}
We use 
the multiplication rule of the $A_4$ triplet  as follow:
\begin{align}
\begin{pmatrix}
a_1\\
a_2\\
a_3
\end{pmatrix}_{\bf 3}
\otimes 
\begin{pmatrix}
b_1\\
b_2\\
b_3
\end{pmatrix}_{\bf 3}
&=\left (a_1b_1+a_2b_3+a_3b_2\right )_{\bf 1} 
\oplus \left (a_3b_3+a_1b_2+a_2b_1\right )_{{\bf 1}'} \nonumber \\
& \oplus \left (a_2b_2+a_1b_3+a_3b_1\right )_{{\bf 1}''} \nonumber \\
&\oplus \frac13
\begin{pmatrix}
2a_1b_1-a_2b_3-a_3b_2 \\
2a_3b_3-a_1b_2-a_2b_1 \\
2a_2b_2-a_1b_3-a_3b_1
\end{pmatrix}_{{\bf 3}}
\oplus \frac12
\begin{pmatrix}
a_2b_3-a_3b_2 \\
a_1b_2-a_2b_1 \\
a_3b_1-a_1b_3
\end{pmatrix}_{{\bf 3}\  } \ , \nonumber \\
\nonumber \\
 {\bf 1} \otimes {\bf 1} = {\bf 1} \ , \qquad &
{\bf 1'} \otimes {\bf 1'} = {\bf 1''} \ , \qquad
{\bf 1''} \otimes {\bf 1''} = {\bf 1'} \ , \qquad
{\bf 1'} \otimes {\bf 1''} = {\bf 1} \  .
\end{align}
More details are shown in the review~\cite{Ishimori:2010au,Ishimori:2012zz}.


\section{Charged lepton and neutrino mass matrices}

The charged lepton mass matrix is given by the multiplication rule of $A_4$ in Appendix B as follows:
\begin{equation}
M_\ell= v_d \alpha_\ell
\begin{pmatrix}
y_e \lambda^4 & 0 & y'_\tau \alpha_{\eta}\\
y'_e \alpha_{\eta}\lambda^4 & y_\mu\lambda^2 &0 \\0& y'_\mu \alpha_{\eta}\lambda^2& y_\tau
\end{pmatrix} \ {\rm for \ \eta\,(1'')} \ , \quad 
v_d \alpha_\ell
\begin{pmatrix}
y_e \lambda^4 & y'_\mu \alpha_{\eta}\lambda^2& 0\\
0 & y_\mu\lambda^2 &y'_\tau\alpha_{\eta} \\y'_e \alpha_{\eta}\lambda^4&0 & y_\tau
\end{pmatrix}   \ {\rm for \ \eta\,(1')} \ ,
\end{equation}
where $\alpha_\ell$, $\alpha_\eta$ and $\lambda$ are  written in terms of the VEVs of  $\phi_T$, $\eta$ and $\Theta$:
\begin{equation}
\alpha_\ell\equiv \frac{\langle \phi_T \rangle}{\Lambda}=\frac{v_T}{\Lambda}\ ,
\qquad
 \alpha_{\eta}\equiv \frac{\langle \eta \rangle}{\Lambda}=
\frac{q}{\Lambda}\ , 
 \qquad  \lambda\equiv \frac{\langle \Theta \rangle}{\Lambda} \ .
\end{equation}
The coefficients $y_i$ and $y'_i$ are order one parameters.

The left-handed mixing matrix of the charged lepton is derived from the diagonalization of  $ U_\ell M_\ell M_\ell^\dagger U_\ell^\dagger$.
The diagonalizing matrix $U_l^\dag$ for the charged lepton is given as follows:
\begin{eqnarray}
&&U_\ell^\dagger \simeq 
\begin{pmatrix}
1 &  -\frac{y'_\mu}{y_\mu} \frac{y'_\tau}{y_\tau} \alpha_{\eta}^2 & \frac{y'_\tau}{y_\tau} \alpha_{\eta}\\
(\frac{y'_\mu}{y_\mu} \frac{y'_\tau}{y_\tau} )^*\alpha_{\eta}^2 & 1& 
\frac{y_\mu y'^*_\mu}{|y_\tau|^2} 
\alpha_{\eta}  \lambda^4 \\
- (\frac{y'_\tau}{y_\tau} )^* \alpha_{\eta}
& \frac{y'_\mu}{y_\mu}  |\frac{y'_\tau}{y_\tau} |^2 \alpha_{\eta}^3 & 1
\end{pmatrix} \quad {\rm for \ \eta\,(1'')} \ ,  \nonumber \\
\nonumber \\
&&U_\ell^\dagger \simeq 
\begin{pmatrix}
1 & \frac{y'_\mu}{y_\mu}\alpha_{\eta} 
& \frac{y'_\tau}{y_\tau} \frac{y'_\mu}{y_\tau}(\frac{y_\mu}{y_\tau})^*\alpha_{\eta}^2 \lambda^4\\
- (\frac{y'_\mu}{y_\mu} )^* \alpha_{\eta}& 1& \frac{y'_\tau}{y_\tau}\alpha_{\eta} \\
(\frac{y'_\mu}{y_\mu}\frac{y'_\tau}{y_\tau})^*\alpha_{\eta}^2&  -(\frac{y'_\tau}{y_\tau})^*\alpha_{\eta} & 1
\end{pmatrix}  \quad {\rm for \ \eta\,(1')} \ .
\end{eqnarray}
The mass eigenvalues of the charged leptons are  given in a good approximation:
\begin{align}
m_e=|y_e|\alpha_\ell \lambda^4 v_d \ ,\quad
m_\mu=|y_\mu|\alpha_\ell \lambda^2 v_d \ , \quad
m_\tau= |y_\tau|\alpha_\ell v_d \ ,
\end{align}
where the Yukawa couplings are of order one.


For $\eta\,(1'')$, the neutrino mass matrix  is given as:
\begin{equation}
M_\nu=a
\begin{pmatrix}
1 & 0 &0 \\
0 & 1 &0 \\0& 0& 1
\end{pmatrix}
+ b
\begin{pmatrix}
1 & 1 &1 \\
1 & 1 &1 \\ 1& 1 & 1
\end{pmatrix}
+c
\begin{pmatrix}
1 & 0 &0 \\
0 & 0 &1 \\
0& 1& 0
\end{pmatrix}
+d
\begin{pmatrix}
0 & 1 &0 \\
1 & 0 & 0 \\ 0& 0& 1
\end{pmatrix} \ ,
\end{equation}
where $a+3b=0$ is satisfied.
  The coefficients $a,b,c$ and $d$ are given in terms of the Yukawa couplings and the VEVs of flavons:
\begin{equation}
a=\frac{y_S\alpha_{\nu}}{\Lambda} v_u^2 \ , \quad
b=-\frac{y_S\alpha_{\nu}}{3\Lambda} v_u^2 \ , \quad
c=\frac{y_\xi\alpha_{\xi}}{\Lambda} v_u^2 \ , \quad 
d=\frac{y'_7\alpha_{\xi}\alpha_{\eta}}{\Lambda} v_u^2 \ , 
\label{abcdA}
\end{equation}
with
\begin{equation}
\alpha_{\nu}\equiv \frac{\langle \phi_S \rangle}{\Lambda}=
\frac{v_S}{\Lambda}\ , \qquad
\alpha_{\xi}\equiv \frac{\langle \xi \rangle}{\Lambda}=
\frac{u}{\Lambda}\ .
\end{equation}
Taking  $a$ to be real  without loss of generality,
we reparametrize them as follows:
\begin{equation}
a\rightarrow a \ , \quad  c\rightarrow c \ e^{i\phi_c} \ , \quad  
d\rightarrow d\  e^{i\phi_d} \ ,
\end{equation}
where $a$, $c$ and $d$ are real parameters and $\phi_c$, $\phi_d$ are CP violating phases.

For $\eta\,(1')$, we get 
\begin{equation}
M_\nu=a
\begin{pmatrix}
1 & 0 &0 \\
0 & 1 &0 \\0& 0& 1
\end{pmatrix}
+ b
\begin{pmatrix}
1 & 1 &1 \\
1 & 1 &1 \\ 1& 1 & 1
\end{pmatrix}
+c
\begin{pmatrix}
1 & 0 &0 \\
0 & 0 &1 \\
0& 1& 0
\end{pmatrix}
+d
\begin{pmatrix}
0 & 0 &1 \\
0 & 1 & 0 \\ 1& 0& 0
\end{pmatrix} \ .
\end{equation}
We obtain the neutrino mass matrix in the TBM basis by rotating it with $V_{\rm TBM}$,
\begin{equation}
V_{\text{TBM}}=
\begin{pmatrix}
\frac{2}{\sqrt{6}} & \frac{1}{\sqrt{3}} & 0 \\
-\frac{1}{\sqrt{6}} & \frac{1}{\sqrt{3}} & -\frac{1}{\sqrt{2}} \\
-\frac{1}{\sqrt{6}} & \frac{1}{\sqrt{3}} & \frac{1}{\sqrt{2}}
\end{pmatrix} \ ,
\end{equation}
as follows: 	
\begin{equation}
\hat M_\nu = V^T_{TBM} M_\nu  V_{TBM}
=
\begin{pmatrix}
a+c e^{i\phi_c}-\frac{d}{2} e^{i\phi_d} & 0 & \mp\frac{\sqrt{3}}{2}d e^{i\phi_d}\\
0 & ce^{i\phi_c}+de^{i\phi_d} &0 \\\mp\frac{\sqrt{3}}{2}d e^{i\phi_d}& 0& a-ce^{i\phi_c}+\frac{d}{2}e^{i\phi_d}
\end{pmatrix},
\end{equation}
where upper (lower) sign  in front of (1,3) and (3,1) components correspond to  
the assignment of $1''$ and $1'$ for $\eta$, respectively.
Next, we consider
\begin{equation}
\hat M_\nu   \hat M_\nu^\dagger = 
\begin{pmatrix}
(1,1) & 0 & (1,3)\\
0& |c e^{i\phi_c}+d e^{i\phi_d}|^2 &0 \\ (1,3)^* & 0& (3,3)
\end{pmatrix},
\end{equation}
where
\begin{eqnarray}
&&(1,1)= a^2+c^2+d^2+2ac\cos\phi_c-cd\cos(\phi_c-\phi_d)-ad\cos\phi_d \ , \nonumber\\
&&  (3,3)= a^2+c^2+d^2-2ac\cos\phi_c-cd\cos(\phi_c-\phi_d)+ad\cos\phi_d \ , \nonumber\\
&&  (1,3)= \mp \sqrt{3}\ [ad\cos\phi_d+icd\sin(\phi_c-\phi_d)]\ .
\end{eqnarray}
We obtain the neutrino mass eigenvalues for NH  as follows:
\begin{eqnarray}
m_1^2&=&a^2+c^2+d^2-cd\cos(\phi_c-\phi_d) \nonumber \\
&&-\sqrt{3c^2d^2\sin^2(\phi_c-\phi_d) +4a^2(c^2\cos^2\phi_c+d^2\cos^2\phi_d-cd\cos\phi_c\cos\phi_d)} \ , \nonumber\\
m_3^2&=&a^2+c^2+d^2-cd\cos(\phi_c-\phi_d) \nonumber \\
&&+\sqrt{3c^2d^2\sin^2(\phi_c-\phi_d) +4a^2(c^2\cos^2\phi_c+d^2\cos^2\phi_d-cd\cos\phi_c\cos\phi_d)} \ , \nonumber\\
m_2^2&=&c^2+d^2+2cd\cos(\phi_c-\phi_d) \ .
\label{masses}
\end{eqnarray}
 The $\hat M_\nu   \hat M_\nu^\dagger$ is diagonalized by the (1-3) family rotation
 as: 
 \begin{equation}
 U_\nu \  (\hat M_\nu   \hat M_\nu^\dagger) \ U_\nu^\dagger = 
 \begin{pmatrix}
 m_1^2 & 0 & 0 \\
 0& m_2^2 &0 \\0 & 0& m_3^2
 \end{pmatrix},
 \end{equation}
 where
  \begin{equation}
U_\nu^\dagger = 
 \begin{pmatrix}
\cos\theta & 0 & \sin\theta e^{-i\sigma}\\
 0& 1&0 \\-\sin\theta e^{i\sigma} & 0& \cos\theta
 \end{pmatrix}.
 \end{equation}
 The $\theta$ and $\sigma$
 are given in terms of parameters in the neutrino  mass matrix:
 \begin{equation}
\tan 2\theta=\sqrt{3}\  \frac{d\sqrt{a^2 \cos^2\phi_d+c^2\sin^2(\phi_c-\phi_d)}}
{ a(d\cos\phi_d-2c\cos\phi_c)} \ , \qquad\quad
 \sigma=-\frac{c\sin(\phi_c-\phi_d)}{a\cos \phi_d} \ .
 \end{equation}
  The parameters $a$, $c$ and $d$ are written in terms of
   $m_1$ and $\alpha_\eta$.
  As seen in Eq.(\ref{abcdA}) , the parameter $d$ is  related with $c$ as
  \begin{equation}
 \frac{d}{c}=\left | \frac{y'_7}{y_\xi} \right | \alpha_\eta\equiv\alpha^\nu_\eta \ ,
 \label{c/d}
 \end{equation}
 where $y'_7$ and $y_\xi$ are of order one coefficients.
 On the other hand, $a$ and $c$ are given in terms of $m_1$, $\alpha^\nu_\eta$,
   $\Delta m^2_{\rm 31}\equiv m_3^2 - m_1^2$ and 
    $\Delta m^2_{\rm 21}\equiv m_3^2 - m_1^2$ 
  since we have the following relations in Eq.(\ref{masses}):
  \begin{align}
\frac{1}{4} (\Delta m^2_{\rm 31})^2&=3c^2d^2\sin^2(\phi_c-\phi_d)
 +4a^2(c^2\cos^2\phi_c+d^2\cos^2\phi_d-cd\cos\phi_c\cos\phi_d) \ ,  \nonumber \\
\Delta m^2_{\rm 21}&= c^2+d^2+2cd\cos(\phi_c-\phi_d) -m_1^2 \ .
  \end{align}
  Then, putting    $\Delta m^2_{\rm atm}=\Delta m^2_{\rm 31}$ and  $\Delta m^2_{\rm sol}=\Delta m^2_{\rm 21}$,
    \begin{eqnarray}
   c^2=\frac{\Delta m^2_{\rm sol}+m_1^2}{1+ (\alpha^\nu_\eta)^2
   +2\alpha^\nu_\eta\cos(\phi_c-\phi_d)} \ ,  \quad
    a^2=\frac{1}{16c^2}\ 
  \frac{\Delta m^2_{\rm atm}-12c^4(\alpha^\nu_\eta)^2\sin^2(\phi_c-\phi_d) }
  {\cos^2\phi_c+(\alpha^\nu_\eta)^2\cos^2\phi_d-\alpha^\nu_\eta\cos\phi_c\cos\phi_d} \ ,
  \end{eqnarray}
  where $m_1$ and $\alpha^\nu_\eta$ are free parameters
  as well as  $\phi_c$ and $\phi_d$.
  
 We comment on the case of  IH of neutrino masses.
 In this case, the neutrino mass eigenvalues are given as
 \begin{eqnarray}
 m_1^2&=&a^2+c^2+d^2-cd\cos(\phi_c-\phi_d) \nonumber \\
 &&+\sqrt{3c^2d^2\sin^2(\phi_c-\phi_d) +4a^2(c^2\cos^2\phi_c+d^2\cos^2\phi_d-cd\cos\phi_c\cos\phi_d)} \ , \nonumber\\
 m_3^2&=&a^2+c^2+d^2-cd\cos(\phi_c-\phi_d) \nonumber \\
 &&-\sqrt{3c^2d^2\sin^2(\phi_c-\phi_d) +4a^2(c^2\cos^2\phi_c+d^2\cos^2\phi_d-cd\cos\phi_c\cos\phi_d)} \ , \nonumber\\
 m_2^2&=&c^2+d^2+2cd\cos(\phi_c-\phi_d) \ ,
 \end{eqnarray}
  where $m_1^2$ and  $m_3^2$ are exchanged each other in Eq.(\ref{masses}).
  Then,  $\Delta m^2_{\rm sol}$ is given as
  \begin{eqnarray}
 \Delta m^2_{\rm sol}&=&m_2^2-m_1^2=
 3cd\cos(\phi_c-\phi_d)-a^2 \nonumber \\
 &&-\sqrt{3c^2d^2\sin^2(\phi_c-\phi_d) +4a^2(c^2\cos^2\phi_c+d^2\cos^2\phi_d-cd\cos\phi_c\cos\phi_d)}  \ .
 \end{eqnarray}
  It is impossible to reproduce the observed value of $\Delta m^2_{\rm sol}$
  since $a\sim c$ and $c\gg d$ in our model as seen in Eq.(\ref{abcdA}).
  Indeed, $d/c$ is  expected  to be $0.1$--$0.2$  in our numerical analysis.  

    Finally, the PMNS matrix is given as 
  \begin{equation}
 U_{\rm PMNS}=U_\ell\  V_{\rm TBM}\  U_\nu^\dagger\  P \ ,
  \end{equation}
  where $P$ is the diagonal phase matrix originated from the Majorana phases.
   The  $P$ is obtained from 
    \begin{equation}
   P U_\nu \hat M_\nu U_\nu^TP={\rm diag}\{m_1, m_2, m_3\},
   \end{equation}
   where $m_1$, $m_2$ and $m_3$ are real positive.
    The effective mass for the $0\nu\beta\beta$ decay is given as follows:
    \begin{align}
    |m_{ee}|=\left|m_1U^2_{e1}+m_2U^2_{e2}+m_3U^2_{e3}\right|  \ .
    \end{align}
  
  \section{Distribution of $\alpha_\eta$}
  The magnitude of the parameter  $\alpha_\ell$ is determined by the tau-lepton mass
  as seen in Eq.(\ref{alpha}).
  The key parameter $\alpha_\eta$ is related with  $\alpha_\ell$
  through  the vacuum structure as discussed in Eq.(\ref{eq:alignment}):
    \begin{align}
   \alpha_\eta=\frac{\lambda _1}{\lambda _2}\sqrt{\frac{g_4}{3g_3}}\ \alpha_\ell \ .
   \label{alphaeta}
   \end{align}
   The coefficients $\lambda_{1(2)}$ and $g_{3(4)}$ are of  order one.
   Then, the factor in front of  $\alpha_\ell$ in Eq.(\ref{alphaeta}) could be
   ${{\cal O}(10)}$.
    We scan $\alpha_\eta$ by using Eq.(\ref{alphaeta})  after fixing  $\alpha_\ell$ in the statistical approach.
    For this purpose, we use the Gamma distribution
    which is available to find the distribution of the order one parameter:
       \begin{align}
   f=(x-\mu)^{(\alpha\gamma-1)} e^{\left ( \frac{x-\mu}{\beta}\right )^\gamma} \ .
   \end{align} 
    Taking  $\gamma=1$  with   $\alpha=3/2$, $\mu=0$ and $\beta=2$,
   we obtain 
   \begin{align}
   f= \sqrt{x}  \ e^{-  \frac{1}{2} x } \ ,
   \label{Gamma1}
   \end{align} 
   which is equivalent to the $\chi^2$ distribution.
   When we take $\gamma=2$ with   $\alpha=1$, $\mu=0$ and $\beta=\sqrt{2}$,
   we obtain 
   \begin{align}
   f= x  \ e^{-  \frac{1}{2} x^2 } \ ,
   \label{Gamma2}
   \end{align} 
   which damps as like the Gaussian distribution at the large $x$.

 It is easy to check that $f$ is maximal at $x=1$ and $f=0$ at $x=0$ for both the two types of Gamma distribution.
 We obtain the distribution of $\alpha_\eta$ by multiplying $\alpha_\ell$ by $f$, which is used in our numerical calculations.
  We show  the distribution of $\alpha_\eta$ in Figs.\,13 and 14
   for $\alpha_\ell=0.0316$  (MSSM $\tan\beta=3$) and $\alpha_\ell=0.010$ (SM)
   in the case of the distributions of Eqs.(\ref{Gamma1}) and (\ref{Gamma2}),
   respectively.

   \begin{figure}[h!]
   \begin{minipage}[]{0.47\linewidth}
    \includegraphics[{width=\linewidth}]{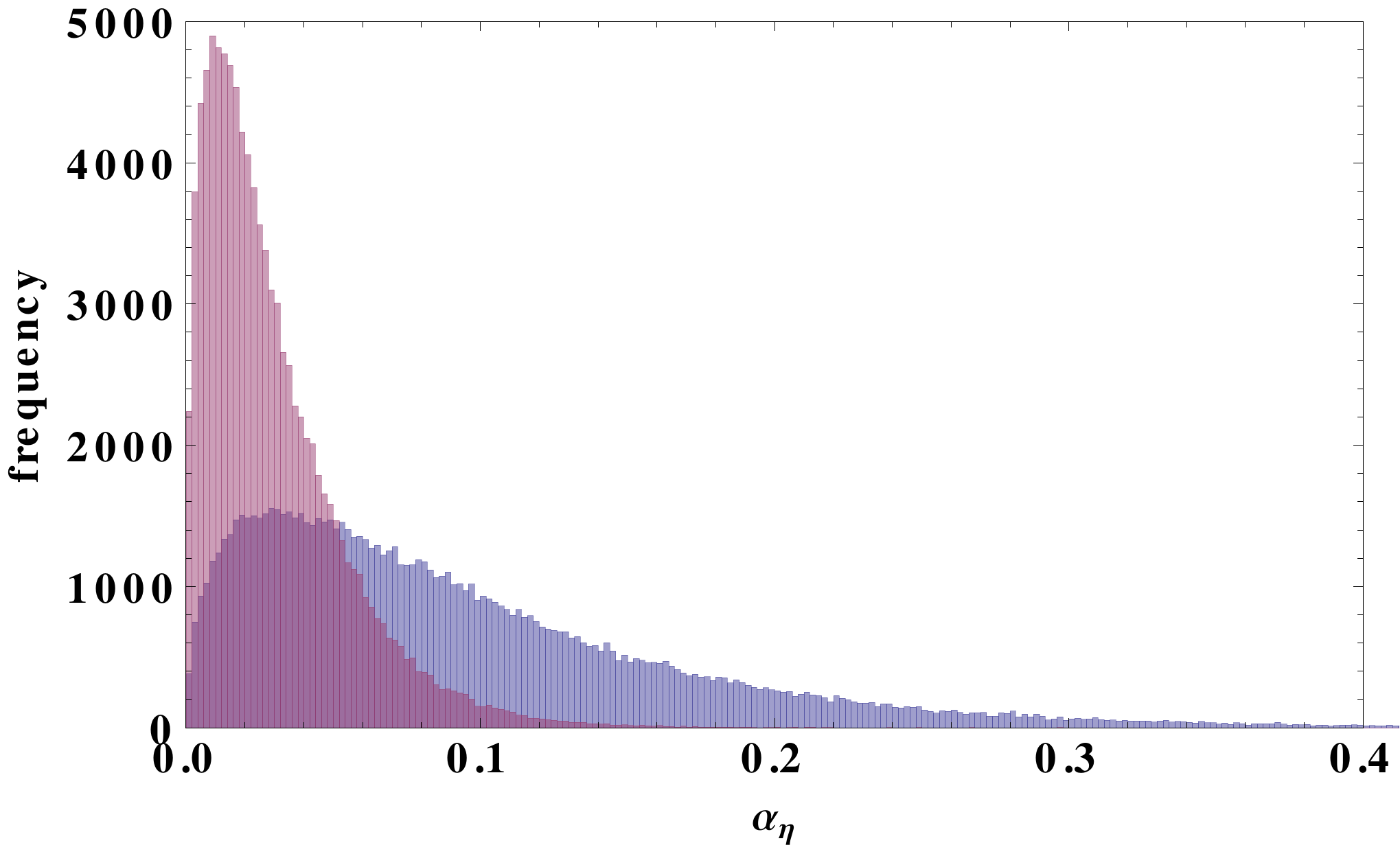}
   \caption{ $\alpha_\eta$ distribution for $\alpha_\ell=0.0316$ (blue) and 
   	 $\alpha_\ell=0.010$ (red)
   	 in  Eq.(\ref{Gamma1}) ($\alpha=3/2, \beta=2, \gamma=1,\mu=0$).}
   \end{minipage}
   \hspace{5mm}
   \begin{minipage}[]{0.47\linewidth}
   \includegraphics[{width=\linewidth}]{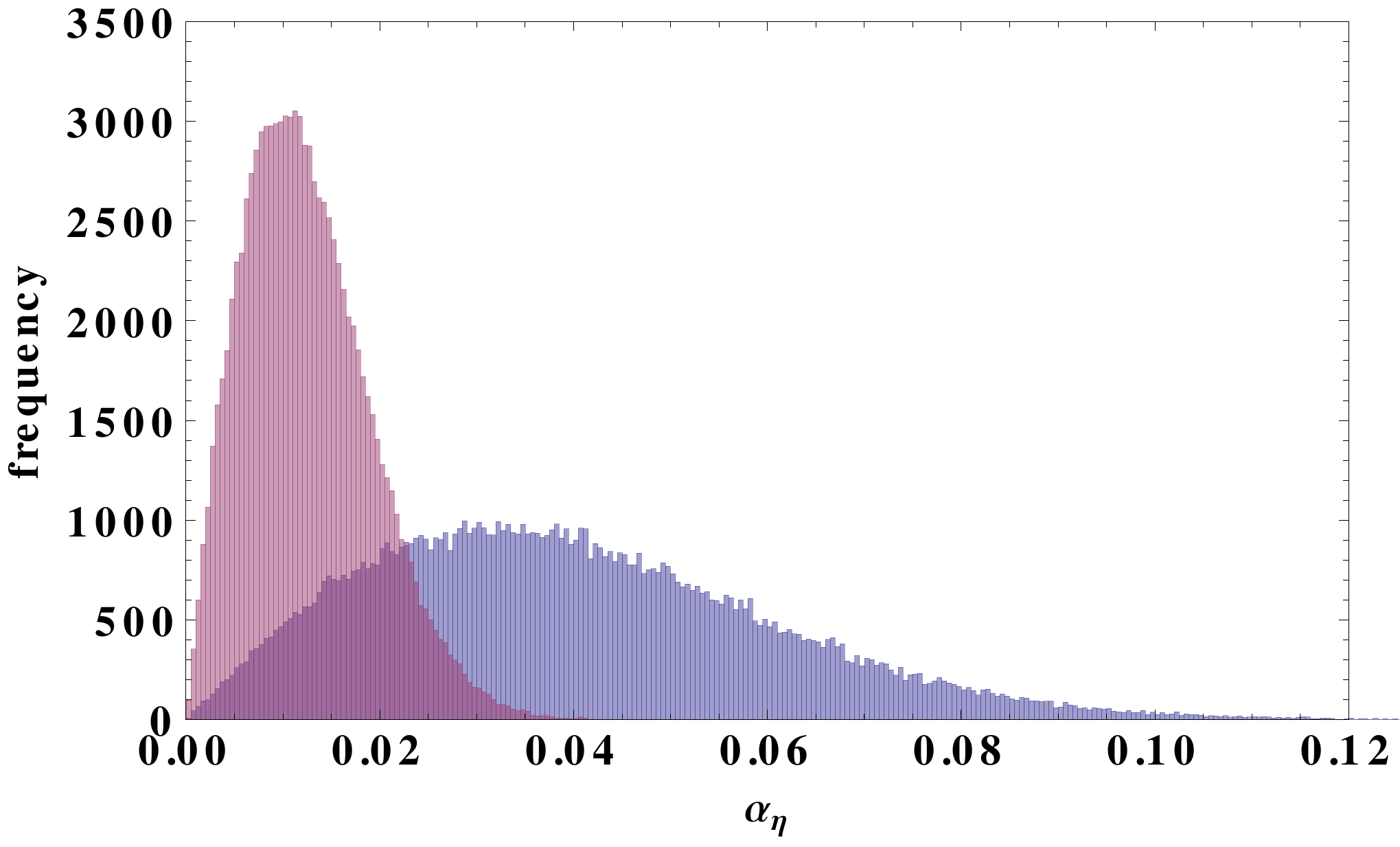}
   \caption{$\alpha_\eta$ distribution for $\alpha_\ell=0.0316$ (blue) and 
   	$\alpha_\ell=0.010$ (red) in Eq.(\ref{Gamma2})
   ($ \alpha=1, \beta=\sqrt{2}, \gamma=2,\mu=0$).}
   \end{minipage}
   \end{figure}


\end{document}